\documentstyle[12pt,epsf]{article}
\begin{document}
\thispagestyle{empty}
\newlength{\defaultparindent}
\setlength{\defaultparindent}{\parindent}

\begin{center}
{\large{\bf Interaction Energies of Generalised Monopoles}}
\vspace{0.5cm}\\

\bigskip
\bigskip
\bigskip
\bigskip
\bigskip

{\large B. Kleihaus,$^{\star}$\footnote{on leave of absence from Fachbereich
Physik, Universit\"at Oldenburg, D-26111 Oldenburg, Germany}
D. O'Keeffe$^{\star}$
and D.H. Tchrakian$^{\star \dagger}$}\\
$^{\star}${\small {\it Department of Mathematical Physics,}\
{\it National University of Ireland, Maynooth,}\
{\it Maynooth, Ireland}\\
$^{\dagger}${\it School of Theoretical Physics,}\
{\it Dublin Institute for Advanced Studies,}\\
{\it 10 Burlington Road,}\
{\it Dublin 4, Ireland.}}\vspace{0.5cm}\\

\end{center}

\bigskip
\bigskip
\bigskip
\bigskip
\begin{abstract}
Generalisations of the 't~Hooft-Polyakov monopole which can exhibit
{\it repulsion} only, {\it attraction} only, and both {\it attraction} and
{\it repulsion}, between like monopoles, are studied numerically. The
models supporting these solitons are $SO(3)$ gauged Higgs models featuring
Skyrme-like terms.
\end{abstract}

\vfill
\setcounter{page}0
\renewcommand{\thefootnote}{\arabic{footnote}}
\setcounter{footnote}0
\newpage

\newcommand{\ra}{\rightarrow}

\pagestyle{plain}

\newcommand{\dd}{\mbox{d}}\newcommand{\tr}{\mbox{tr}}
\newcommand{\ee}{\end{equation}}
\newcommand{\be}{\begin{equation}}
\newcommand{\ii}{\mbox{i}}\newcommand{\e}{\mbox{e}}
\newcommand{\pa}{\partial}\newcommand{\Om}{\Omega}
\newcommand{\vep}{\varepsilon}
\newcommand{\bfph}{{\bf \phi}}
\newcommand{\lm}{\lambda}
\def\theequation{\arabic{equation}}
\renewcommand{\thefootnote}{\fnsymbol{footnote}}
\newcommand{\re}[1]{(\ref{#1})}
\newcommand{\bfR}{{\sf R\hspace*{-0.9ex}\rule{0.15ex}%
{1.5ex}\hspace*{0.9ex}}}
\newcommand{\N}{{\sf N\hspace*{-1.0ex}\rule{0.15ex}%
{1.3ex}\hspace*{1.0ex}}}
\newcommand{\Q}{{\sf Q\hspace*{-1.1ex}\rule{0.15ex}%
{1.5ex}\hspace*{1.1ex}}}
\newcommand{\C}{{\sf C\hspace*{-0.9ex}\rule{0.15ex}%
{1.3ex}\hspace*{0.9ex}}}
\newcommand{\1}{{\sf 1\hspace*{-0.9ex}\rule{0.15ex}%
{1.3ex}\hspace*{0.9ex}}}
\renewcommand{\thefootnote}{\arabic{footnote}}

\section{Introduction}
\setcounter{equation}{0}

It is well known that 't Hooft-Polyakov \cite{tH,P} monopoles 
of like magnetic charge repel.
For all finite values of the Higgs mass, or the coupling constant of the Higgs
self-interaction potential,
like monopoles exhibit a single repulsive phase only. 
This is understood as follows:
In the Prasad-Sommerfield (PS) \cite{PS} limit, the Higgs field
is massless and hence mediates a long range attractive force that
cancels the long range repulsive magnetic force of the $U(1)$ field, exactly.
In this limit, all components of the stress tensor vanish~\cite{JT} resulting
in the vanishing of inter-monopole forces.
The force between like monopoles in the BPS limit was studied in detail by
Manton \cite{M} and Nahm \cite{N}, who showed
that it decreases faster than any inverse power. 
In the presence of the usual Higgs potential however,
the Higgs field becomes massive and as a result decays exponentially.
Consequently the long range magnetic field dominates at large distances,
leading to the repulsion of like monopoles of the Georgi-Glashow (GG) model. 
This was also concluded by Goldberg et al.~\cite{Gold}, 
using the time-rate of change
of the stress tensor for the field configuration of two exponetially
localized monopoles, 
situated apart at a distance much larger than the
sizes of the monopole cores. 
Recently Kiselev and Shnir~\cite{KS}
also have shown this employing the fluctuations and zero modes
around the monopoles. 
That there is only one repulsive phase has been
verified in \cite{KKT}, by calculating the energy 
{\it per unit topological charge}
for a charge-2 and charge-1 solution numerically for all values
of the Higgs potential coupling constant. 
The procedure in Ref.~\cite{KKT}
is exactly analogous to that adopted by Jacobs and Rebbi~\cite{JR} in their
investigation of attractive and repulsive phases in the Abelian Higgs model.

Implementing the procedure of Ref.~\cite{JR} in our 3 dimensional context
necessitates the numerical solution of the partial differential equations
of the axially symmetric fields. 
This is because the charge-2 monopole
in 3 dimensions is not spherically symmetric,
unlike in the case of the 2 dimensional context of the Abelian Higgs
model~\cite{JR}
and of the hierarchy of Abelian Higgs models~\cite{ABT}, 
where vorticity-$n$ solutions are radially symmetric 
and hence result from ordinary differential equations. 
Thus, one of the most important technical ingredients of this work
is the imposition of the axial symmetry on the $SO(3)$ gauge field and the
algebra valued Higgs, 
for which we use the Ansatz of Rebbi and Rossi~\cite{RR}.

The purpose of the present work is to show 
that certain generalisations of the monopoles of the GG model, 
exhibit 
(a) a {\it repulsive} phase only like the usual monopole, 
(b) an {\it attractive} only, and 
(c) both {\it attractive} and {\it repulsive} phases. 
This is done by integrating the Euler-Lagrange equations arising from
the static Hamiltonians of the new models, numerically, 
in exact analogy with the
procedures used in Refs.~\cite{JR,KKT}. 
The models under consideration, 
just like the GG model, 
are 3 dimensional $SO(3)$ Higgs models. 
These models are
constructed from the hierarchy of 3 dimensional $SO(3)$ Higgs models, 
obtained 
by dimensional descent from the $4p$ dimensional hierarchy of chiral
$SO(4p)$ Yang-Mills models~\cite{T}
\footnote{There are similar hierarchies of $d$ dimensional $SO(d)$ Higgs 
models with $d<4p$~\cite{TC}}. 
Here, we will restrict to the $p=1$ and the
$p=2$ members of this hierarchy, of which the $p=1$ member is nothing else 
than the PS limit of the GG model. 
Thus the main new ingredient in the models presented here
will be the $p=2$ model, whose spherically symmetric monopol solutions
were studied numerically in our previous work~\cite{KOT}, 
to which we shall refer for certain details later.

In Ref.~\cite{KOT}, we have discussed the 
charge-1 solutions of the $p=2$ model in some
detail, especially with regard to its properties relative to the GG and
the Abelian Higgs models. There we have pointed out that the $p=2$ model
does not support self-dual solutions. 
However, for certain values of the coupling constants
it nevertheless supports a
solution whose energy is quantitatively
extremely close to the Bogomol'nyi lower bound, and
correspondingly, an estimate of some components of the stress tensor 
indicates that the force on the soliton is also extremely small. 
We described such a
solution as being ``almost self-dual" and
we designated the corresponding values of the coupling
constants as {\it critical}. 
This is in accordance with the
usual nomenclature whereby the Bogomol'nyi inequalities are saturated for
the {\it critical} values of the coupling constants in a given
model. 
It is important to recall~\cite{TC,KOT} that the {\it critical} 
dimensionless coupling constants of the $p=2$ model take 
{\it nonzero} values.
In this sense the $p=2$
model is similar to the Abelian Higgs model,
which supports a self-dual solution for a finite value of the 
dimensionless coupling constant, and it is in contrast to
the GG model for which the critical value of the (Higgs)
coupling constant {\it vanishes}. 
Hence it was argued in Ref.~\cite{KOT} that
like the Abelian Higgs model, the 3 dimensional $p=2$ $SO(3)$ Higgs model,
supports both attractive and repulsive phases. 

To put all this in perpective, we note that there is a difference between
the $p=2$ model on the one hand and the $p=1$ and the abelian Higgs model
on the other hand.
In the latter, the Bogolmol'nyi inequality is saturated at the 
{\it critical}  coupling constants for all topological charges $n$.
Thus the {\it critical} values of the coupling constants characterise 
a property of the model itself. For the $p=2$ model we only know that 
for the charge-1 solution the energy is close to the Bogolmol'nyi bound
at the {\it critical} coupling constants. If these coupling constants 
characterise the model, and not only a single solution, then the 
energy of all charge-$n$ solutions should be near their  Bogolmol'nyi 
bounds at the same values of the coupling constants.
So we do not specify quantatively what we mean by `near the Bogolmol'nyi 
bound', there might be a subset of values of the coupling constants 
which can be called {\it critical}. For the charge-2 solution of the 
$p=2$ model we expect that the energy will be near the  Bogolmol'nyi bound
for coupling constants close to the {\it critical} coupling constants 
determined by the investigation of the charge-1 solutions.
If this is the case, attractive and repulsive phases possibly exist with
a crossover at values of the coupling constants near the {\it critical} 
values.

It is one of the aims of the present work to verify this hypothesis.
This will be done by
straighforwardly integrating the Euler-Lagrange equations numerically
on the charge-1 and charge-2 sectors. 
Thus the argument based on the {\it almost self-duality} of the $p=2$
monopoles, that makes the existence of the two phases plausible, 
will not be further elaborated here, 
and we refer to Refs.~\cite{ABT,KOT} for that.

In general, our aim is to present all the generalised $SO(2)$ Higgs models
with Hamiltonian densities carrying dimensions up to $L^{-8}$, which includes
of course the GG model with dimension $L^{-4}$. Note that the limitation we
impose is a consequence of limiting our study to the $p=2$ model,
whose Hamiltonian density has the dimensions $L^{-8}$. The inclusion of the
GG or the $p=1$ model results in the presence of the usual Yang-Mills term
and hence in the possibility of describing free gauge fields, absent in the
pure $p=2$ (and $p\ge 2$) model(s). We note that the $p\ge 2$ models are 
purely Skyrme-like gauged Higgs systems~\cite{T2} 
in the sense that they consist of
$2p$th powers of the curvature and covariant derivative, suitably
antisymmetrised
so that only the squares of velocity-fields appear. In this sense, the
composite models consisting of the $p=1$ model augmented by $p\ge 2$ models,
are quite analogous to the Skyrme model~\cite{S} where the latter terms play
the role of the Skyrme term of the $O(4)$ sigma model.

It is worthwhile making some clarifications concerning the 
construction of the composite models. 
We know from the results and the arguments of
Refs.~\cite{ABT,KOT} that we would expect a phase change from attractive
to repulsive, to take place at the {\it almost self-dual} configurations of
the systems.
The composite models however are even less self-dual than the 
$p=1$ and the $p=2$ modes on their own, since the Bogolmol'nyi equations
are even more overdetermined then those of the $p=2$ model on its own
and hence the solutions of the composite models are more remote from
saturating the Bogolmol'nyi bound.
(Clearly,  exact self-dual solutions can 
be achieved if we suppress the entire $p=2$ content of the model,
but then there will be no interactions between the monopoles. 
Since our aim is to produce a (composite) model accomodating interactions,
we will resist on retaining some or all of the components of the
$p=2$ model.)
We will investigate 
three possibilities, (a) when the $p=1$ system dominates, which can be
regarded as a $p=1$ model extended with a $p=2$ Skyrme-term, (b) when the
two constituents of the composite model are equally important, and (c)
when the $p=2$ model dominates, which yields the pure $p=2$ model whose
spherically symmetric solutions were studied in Ref.~\cite{KOT}. 
We shall study all these three cases in this work.

In Section 2 we define the static Hamiltonians of the various models we will
study and proceed in Section 3 to state the axially symmetric Ansatz of Rebbi
and Rossi~\cite{RR}, leading to the residual 2 dimensional subsytems of the
models
introduced in the previous section. In Sections 4, 5 and 6, we present the
results of
the numerical computations, respectively for the extended $p=1$ models where
the latter dominates, the composite $p=1$ and $p=2$ system, and the pure
$p=2$ system which is the limiting case of the composite system where the 
$p=2$
dominates. Our choice for this order of presentation is the result of the
successively increasing complication involved in the numerical computations
involved in each of these cases, so that we have chosen to tackle the most
easily accessible cases first. Section 7, finally, is devoted to the summary
and discussion of the results.

\section{The Models}

Since various aspects of the 3 dimensional hierarchy of $SO(3)$ gauged
Higgs models are discussed in detail in Refs.~\cite{TC,KOT,T2}, we shall
restrict ourselves here to the definitions of the Hamiltonian densities
only. We refer to Refs.~\cite{TC,T2} for their construction, as well as the
topological inequalities, via dimensional descent. For the detailed
normalisation of the topological charge and the asymptotic properties
required by the finite energy condition and analyticity,
we refer to Ref.~\cite{KOT}.

Below, we shall denote the gauge and Higgs fields
\be
\label{0}
A_i =-{i\over 2}A_i^a \sigma_a \: , \qquad \phi = -{i\over 2}\phi^a \sigma_a
\ee
in an antihermitian notation. The resulting antihermitian curvature and
covariant
derivative are then
\be
\label{c}
F_{ij} =\pa_i A_j -\pa_j A_i +[A_i ,A_j]\: ,
\ee
\be
\label{cd}
D_i \phi =\pa_i \phi +[A_i ,\phi].
\ee

\subsection{The $p=1$ model}

This is the PS limit of the GG model, characterised by the static Hamiltonian
\be
\label{1}
{\cal H}^{(1)} = \mbox{Tr} ({1\over 4}F_{ij}F^{ij} +{1\over 2} 
D_i \phi D^i \phi) \ .
\ee
The reason that we have distinguished between covariant and contravariant
indices is that in what follows we shall have occasion to use a spherical
(not Cartesian)  basis.

The system supports self-dual solutions saturating the topological
lower bound~\cite{B},
which equals the magnetic charge. This model is extensively studied 
since a long
time and will not be considered in isolation here.

\subsection{The $p=2$ model}

The static Hamiltonian of this model is~\cite{KOT}
\be
\label{2}
{\cal H}^{(2)}=\frac{5}{18\times 32}\sum_{A=1}^{4} \lambda_A {\cal H}_A^{(2)},
\ee
where each of the four densities ${\cal H}_A^{(2)}$ are defined by
\begin{eqnarray}
\label{2.1}
{\cal H}_4^{(2)}
&=& \mbox{Tr}  \{ F_{[ij},D_{k]} \phi \} \{ F^{[ij},D^{k]} \phi \} \ , \\
\label{2.2}
{\cal H}_3^{(2)}
&=& -6\mbox{Tr}  (\{ S,F_{ij} \} +[D_i \phi ,D_j \phi])
(\{ S,F^{ij} \} +[D^i \phi ,D^j \phi])  \ ,  \\
\label{2.3}
{\cal H}_2^{(2)}
&=& -27\mbox{Tr} [ \{ S,D_i \phi \} \{ S,D^i \phi \} ] \ ,\\
\label{2.4}
{\cal H}_1^{(2)}
&=& 54\mbox{Tr} S^4  .
\end{eqnarray}
In \re{2.1}-\re{2.3}, we have used $[ijk]$ to denote cyclic symmetry 
in the indices
$i,j,k$, $\{ A,B\}$ denotes an anticommutator, and
\be
\label{7}
S =  -(\eta^2 +\phi^2).
\ee
The reason that we have distinguished between covariant and contravariant
indices
in \re{2.1}-\re{2.4} is the same as in \re{1}.

In \re{2} we can set $\lambda_4 =1$ by scaling with an overall constant.
($\lambda_4$ is reserved for use only when we construct a composite model by
combining the $p=1$ and $p=2$ models.)
When the dimensionless coupling constants 
$\lambda_1 ,\lambda_2$ and $\lambda_3$
are restricted to their critical values
$\lambda_1 =\lambda_2$ = $\lambda_3 =1$, then
the topological inequalitiy
\be
\label{8}
\int {\cal H}^{(2)}(\lambda_1 =\lambda_2 =\lambda_3 =1)d^3r \ge 4\pi n \eta^5,
\ee
where $n$ is the topological charge,
can be saturated by the corresponding Bogomol'nyi
equations. The latter being overdetermined~\cite{TC} they do not support
nontrivial
solutions. Nevertheless there exist nontrivial solutions, that do not satisfy
the Bogomol'nyi equations, with energies lying above but numerically
very close to the Bogomol'nyi lower bound. The spherically symmetric
solution found in Ref.~\cite{KOT} is of this type, described as being {\it
almost
self-dual}. When all the coupling constants are greater than 1, i.e. 
$\lambda_A
\ge 1$, then the topological inequality \re{8} is preserved. 
When however any of
the coupling constants takes a lower value than 1, i.e. $\lambda_A \le 1$, 
then
the inequality \re{8} is replaced by
\be
\label{9}
\int {\cal H}^{(2)} d^3r \ge 4\pi \lambda_{min} n \eta^5,
\ee
where $\lambda_{min}$ is the smallest of $\lambda_A$, $A=1,2,3$.

>From the results in Ref.~\cite{ABT}, where we have seen that there is a
crossover
point between attractive and repulsive phases in the vicinity of the critical
values of the dimensionless coupling constants of a system supporting
{\it almost self-dual} solutions, we expect this to be the case also for the
$p=2$ model \re{2}, which we know from the results of Ref.~\cite{KOT} that it
does support {\it almost self-dual} solutions.

\subsection{The composite model}

By composite models we mean the system characterised by the sum of the
static Hamiltonians ${\cal H}^{(p)}$, always including the $p=1$ member
of the hierarchy. Here we are limiting ourselves to
$p=1$ and $p=2$. In {\bf 2.3.1}, we define the models we shall
subsequently study numerically. In {\bf 2.3.2}, we estimate the
dependence of the energy on the coupling constants $\lambda_A$ in the region
$\lambda_A \ll 1$.

\subsubsection{The generic case}

The Hamiltonian density of this model is
\be
\label{10}
{\cal H}^{(c)}={\cal H}^{(1)} +\eta^{-4}{\cal H}^{(2)},
\ee
whose Bogomol'nyi lower bound is twice $4\pi n\eta$ if $\lambda_A \ge 1$.

In \re{10}, the dimensionless coupling constants $\lambda_A$ for all four
$A=1,2,3,4$, can take on any positive definite values, 
and even when any or all $\lambda_A$ vanish, 
the Bogomol'nyi lower bound does not sink to zero as for 
the $p=2$ model in \re{9}, 
since the energy of the $p=1$ model is still bounded
by the usual magnetic charge. 
The fact that it is possible to set all but one
of the $\lambda_A$ equal to zero in \re{10} leads us to describe
\re{10} as an extended $p=1$ model.
For example if we set all but one $\lambda_A$ in \re{10}
equal to zero we have four special composite models, 
and if we set any two of
them equal to zero we have six special composite models, etc.
The first case is quite special. These four models are similar to
the GG model in sofar as they describe a $p=1$ model with only one additional
term. This suggests that like the
GG model, these four composite models support only one phase, i.~e.
an attractive
only or a repulsive one only. We shall give a strongly suggestive analytic
argument for this in {\bf 2.3.2} below.

Since the Bogolmol'nyi bounds of the solutions to (\ref{10}) are more
substantially exceeded than in the $p=2$ model, here we cannot invoke the 
presence of almost self-dual solutions. Thus in the generic case, where
$\lambda_A$ are not considered to be small we cannot predict anything
about the interaction ot the monopoles. The situation is different 
however, when all $\lambda_A \ll 1$. In these cases it is possible to 
examine this question systematically as we shall do in the 
following subsections.

In order to determine whether there is a repulsive phase,  
an attractive phase or a crossover we follow Jacobs and Rebbi~\cite{JR},
and define the difference of the 
energy {\it per unit topolgical charge} for the charge-2 and 
charge-1 solutions at fixed coupling constants $\lambda_A$,
\be
\label{16}
\Delta E(\lambda_A) \stackrel{def.}
= \frac{E^{(2)}(\lambda_A)}{2} -E^{(1)}(\lambda_A) \ ,
\ee
where $E^{(n)}(\lambda_A)$ is the energy of the charge-$n$ solution.
The monopoles repel when $\Delta E(\lambda_A) >0$ and
attract when  
$\Delta E(\lambda_A)<0$.
If the quantity $\Delta E(\lambda_A)$ vanishes,
we find a crossover at $\lambda_A$.

\subsubsection{Perturbed $p=1$ models}

What is very interesting about the extended models is, that
the behaviour of the solutions on the constants $\lambda_A$ 
can be analytically
estimated for small values of the coupling constants, $\lambda_A \ll 1$. 
We shall restrict, at first, to the case where there is only one nonzero
$\lambda_A$ in ${\cal H}^{(2)}$.

Consider the energy integral
\be
\label{11}
E^{(n)}(\lambda_A) =\int ({\cal H}^{(1)}
+\frac{5}{18\times 32}\eta^{-4} \lambda_A {\cal H}_A^{(2)}) d^3 x ,
\ee
as a function of $\lambda_A$. The energy density functional in \re{11} is the
special case of the composite static Hamiltonian \re{10} in which all but
a given coupling constant $\lambda_A$ are set equal to zero, and is evatuated
for the topological charge-$n$
solution described by the fields $(A_i^{(\lambda_A)} , \phi^{(\lambda_A)})$.

We now insist that $\lambda_A \ll 1$, and make a Taylor expansion of
$E^{(n)}(\lambda_A)$,
\be
\label{12}
E^{(n)}(\lambda_A) = E^{(n)}(0)
+\lambda_A \left. \frac{dE^{(n)}}{d\lambda_A}\right|_{\lambda_A =0}
+o(\lambda_A^2) .
\ee
Because $\lambda_A \ll 1$, we can describe the solution fields in the energy
density functional in \re{11} as small perturbations of the known
charge-$n$ PS~\cite{PS} solution fields $(A_i^{(0)} ,\phi^{(0)})$
\be
\label{13}
A_i^{(\lambda_A)} =A_i^{(0)} +\lambda_A a_i ,\qquad
\phi^{(\lambda_A)} =\phi^{(0)} +\lambda_A \varphi
\ee
in terms of the functions $(a_i ,\varphi)$ which in principle can be found.

The first term in the Taylor expansion \re{12} is recognised as the energy of
the PS solution, and equals $4\pi n\eta$. To calculate the second term, we
differentiate the energy integral \re{11}, which involves the differentiation
of the functional in the integrand both with respect to its implicit
dependence on $\lambda_A$ through the fields $A_i^{(\lambda_A)},
\phi^{(\lambda_A)}$, and its explicit dependence. The terms arising from the
differentiation of the implicit dependence on $\lambda_A$ amount to subjecting
the functional 
${\cal H}^{(1)} +\frac{5}{18\times 32}\eta^{-4}\lambda_A {\cal H}^{(2)}$ 
to the variational
principle, which vanish by virtue of the Euler-Lagrange equations satisfied 
by the solution fields \re{13}. 
The result of the differentiation with respect to
the explicit dependence on $\lambda_A$ then is the only contribution
\be
\label{14}
\frac{dE^{(n)}}{d\lambda_A}=
\frac{5}{18\times 32}\eta^{-4}\int {\cal H}_A^{(2)}[A_i^{(\lambda_A)},
\phi^{(\lambda)}] d^3 x ,
\ee
which evaluated at $\lambda_A =0$ is
\be
\label{15}
\left. \frac{dE^{(n)}}{d\lambda_A}\right|_{\lambda_A =0}
=\frac{5}{18\times 32}\eta^{-4}\int{\cal H}_A^{(2)}[A_i^{(0)},\phi^{(0)}] 
d^3 x ,
\ee
an integral evaluated strictly in terms of the PS solution fields
$(A_i^{(0)} ,\phi^{(0)})$. This is the sense in which the present calculation
is described as being {\it perturbative}.

Refining our notation for the PS solution fields $(A_i^{(0)} ,\phi^{(0)})$ by
labeling them further with the topological charge $n$ as
$(A_i^{(0)}(n) ,\phi^{(0)}(n))$, we substitute \re{15} in \re{16}
\begin{eqnarray}
\label{18}
\Delta E &=& c_A \lambda_A  \\
c_A &=& \frac{5}{18\times 32}\eta^{-4} \bigg( {1\over 2}
\int {\cal H}_A^{(2)}[A_i^{(0)}(2), \phi^{(0)}(2)] d^3 x
-\int {\cal H}_A^{(2)}[A_i^{(0)}(1), \phi^{(0)}(1)] d^3 x \bigg)  .
\end{eqnarray}
Clearly, when the quantity $c_A$ in \re{18} is positive (negative) the 
monopoles repel (attract). 
Moreover, if we assume that the dependence of the energies on
$\lambda_A$ is monotonic, then this will be the only phase, like in the GG
model. It is important however to stress that the above arguments do not apply
to the GG model, namely, that the GG model cannot be treated as a perturbation
of the $p=1$ (BPS) system \cite{KKT}. 

In table 1 the values of 
$\left.\frac{1}{n} \frac{dE^{(n)}}{d\lambda_A}\right|_{\lambda_A =0}$
for the charge-1 and the charge-2 solutions and $c_A$ are given.
Note, $c_A$, $A=2,3,4$ are of the same order of magnitude, whereas
$c_1$ is more than one order of magnitude larger. 
$c_1$ and $c_2$ are positive, leading to repulsion of like monopoles for 
the models with $\lambda_1 \neq 0$ or $\lambda_2 \neq 0$. 
For the models with $\lambda_3 \neq 0$ or $\lambda_3 \neq 0$ the 
constants $c_3$ and $c_4$ are negative. Hence these models have only 
attractive phases.

The above perturbative argument can be systematically extended to the case of
the full composite model with all or some of the $\lambda_A \not = 0$. 
The only
difference there will be that the Taylor expansion will have more than one
contribution (one from each nonzero $\lambda_A$),
\begin{eqnarray}
\label{18b}
\Delta E &=& \sum_{A=1}^4 c_A \lambda_A  \ . \nonumber\\
\end{eqnarray}
Indeed, it would be expected that there should be both phases, like in
the generic case.
If at least two of the constants $c_A$ have opposite signs then
the model has attractive and repulsive phases. The change of 
the phases occurs for the values of the parameters $\lambda_A$ for which 
$\Delta E$ becomes zero.
This is verified by our numerical studies for small values
of $\lambda_A$, reported later.

\section{Axial Symmetry}

Imposition of axial symmetry proceeds readily by the application of the
Rebbi-Rossi Ansatz~\cite{RR}. When subjected to axial symmetry, the 12
fields consisting of the 3 components of the isospin-1 gauge field and
the isospin-1 Higgs field are parametrised by 6 functions. It is
instructive to describe~\cite{M2,RR} these 6 funcions as a 2 component real
field ${\cal A}_{\mu}$, with $\mu =1,2$ labeling the 2 dimensional
coordinate $\xi_{\mu}= r,z, \qquad r=\sqrt{x^2 +y^2}$, and two complex scalar
fields $\varphi$ and $\chi$. In this parametrisation, all Lorentz and
$SU(2)$ gauge invariant Hamiltonian densities ${\cal H}^{{(p)}}$ reduce
to an Abelian Higgs model in which ${\cal A}_{\mu}$ plays the role of a $U(1)$
connection, and the two complex scalar fields $\varphi$ and $\chi$ play the
roles of Higgs fields. The asymptotic behaviours required of these functions
$\varphi$ and $\chi$ by the requirements of finite energy and analyticity are
typical of Higgs fields, hence justifying the description of the axially
symmetric 2 dimensional reduced subsytems of the $p$-th $SO(3)$ gauged Higgs
model in 3 dimensions, as residual Abelian two-Higgs models in 2 dimensions.
(In particular, one of the complex fields, say $\varphi$ actually is endowed
with a Higgs type self-intereaction potential, and the two Higgs fields
exhibit a Yukawa-type mutual interaction term.) Given that the residual
subsytem exhibits a $U(1)$ gauge symmetry, it is necessary to remove one of
the 6 residual fields by fixing this gauge. The natural choice we will make
here,
which has been used previously in Refs.~\cite{ KK,BK}, is
\be
\label{19}
\pa_{\mu} {\cal A}_{\mu} =0 ,
\ee
which we call a Coulomb gauge as before~\cite{ KK,BK}.

In the following, we shall not describe the axially symmetric subsystem 
in terms
of the functions $({\cal A}_{\mu},\varphi ,\chi)$, but will use rather another
parametrisation formulated in spherical coordinates. This choice is made
because it enables the efficient imposition of the boundary conditions in the
numerical computations to follow.

First we define the $su(2)$ matrices $\tau_{a}^{(n)}$ in terms of 
Pauli matrices $\sigma_a$ and spatial coordiates $\theta$, $\varphi$
\begin{equation}
\tau_{r}^{(n)}      = \sin \theta \cos n\varphi \ \sigma_1 
                     +\sin \theta \sin n\varphi \ \sigma_2
                     +\cos \theta \ \sigma_3  \ ,
%\nonumber
\end{equation}
\begin{equation}
\tau_{\theta}^{(n)} = \cos \theta \cos n\varphi \ \sigma_1 
                     +\cos \theta \sin n\varphi \ \sigma_2
                     -\sin \theta \ \sigma_3  \ ,                
%\nonumber
\end{equation}
\begin{equation}
\tau_{\varphi}^{(n)}   =  -\sin n\varphi \ \sigma_1 
                       +\cos n\varphi \ \sigma_2 \ ,
%\nonumber
\end{equation}
where $n$ is the winding number.
For the Ans\"atze of the gauge field and the Higgs field we 
impose 
axial symmetry and in addition parity reflection symmetry.
Using spherical coordinates the Ansatz of the gauge field 
\begin{equation}
A_i dx^i = A_r dr + A_\theta d\theta + A_\varphi d\varphi
\end{equation}
can be parametrized by  four functions $H_k(r,\theta)$,
\begin{equation}
A_r = \frac{H_1(r,\theta)}{r} \ \frac{i \tau_{\varphi}^{(n)}}{2}\ , \ \ 
A_\theta = (1 - H_2(r,\theta)) \ \frac{i \tau_{\varphi}^{(n)}}{2}\ ,  
%\nonumber
\end{equation}
\begin{equation}
A_\varphi = -n \sin \theta\ ( H_3(r,\theta) \ \frac{i \tau_{r}^{(n)}}{2}
               +(1 - H_4(r,\theta)) \ \frac{i \tau_{\theta}^{(n)}}{2}) \ . \          
\end{equation}
The Ansatz for
the Higgs field can be parametrised by two functions $\phi_k(r,\theta)$
\begin{equation}
\frac{1}{2}\phi = \phi_1(r,\theta) \ \frac{i \tau_{r}^{(n)}}{2} 
     + \phi_2(r,\theta) \ \frac{i \tau_{\theta}^{(n)}}{2} \ .
\end{equation}

Expanding the curvature tensor $F_{ij}$ in terms of the matrices 
$i\tau_{a}^{(n)}/2$,
\begin{equation}
F_{ij} = F^{(a)}_{ij} \ \frac{i \tau_{a}^{(n)}}{2} \ ,
\end{equation}
we find for the non-vanishing coefficients $F^{(a)}_{ij}$,
\begin{eqnarray}
F^{(\varphi )}_{r\theta}& = &
-\frac{1}{r} (r \partial_r H_2 + \partial_\theta H_1)\ ,\\
F^{(r)}_{r\varphi}& = &
    -\frac{n}{r} \sin \theta (r \partial_r H_3 - H_1 \ H_4 \ )
\ , \\
F^{(\theta)}_{r\varphi} &=& 
 \frac{n}{r} \sin \theta (r \partial_r H_4 + H_1 \ (H_3 +\cot \theta ))\ ,\\
F^{(r)}_{\theta \varphi} &=& 
    -n \sin \theta ( \partial_\theta H_3 + H_3 \cot \theta + H_2 \ H_4 -1 )
\ , \\
F^{(\theta)}_{\theta \varphi}& = &
 n \sin\theta (\partial_\theta H_4 + (H_4 - H_2) \cot\theta  - H_2 \ H_3 )\ .
\end{eqnarray}
Similarly we find for the expansion of the covariant derivative of 
the Higgs field,
\begin{equation}
D_\mu \phi = (D_\mu \phi)^{(a)} \ \frac{i \tau_{a}^{(n)}}{2}\ ,
\end{equation}
the non-vanishing coefficients
\begin{eqnarray}
(D_r \phi)^{(r)} &=& \frac{1}{r} (r \partial_r \phi_1 + H_1 \ \phi_2)
\ , \\
(D_r \phi)^{(\theta)} &=& \frac{1}{r} (r \partial_r \phi_2 - H_1 \ \phi_1)
\ , \\
(D_\theta \phi)^{(r)} &=& \partial_\theta \phi_1 - H_2 \ \phi_2
\ , \\
(D_\theta \phi)^{(\theta)} &=& \partial_\theta \phi_2 + H_2 \ \phi_1 \ , \\
(D_\varphi \phi)^{(\varphi)} &=&
n \sin \theta (  \phi_1 \ H_4 + \phi_2 (H_3 + \cot \theta )) \ .
\end{eqnarray}

With these Ans\"atze the Hamiltonian densities eqs.~\re{2.1}-\re{2.4} 
can be expressed as
\begin{eqnarray}
{\cal H}_4^{(2)} &=& 3\left[
            F^{(\varphi )}_{r\theta}(D_\varphi \phi)^{(\varphi)}
            \right. \nonumber\\
            & & +F^{(r)}_{\varphi r}(D_\theta \phi)^{(r)}
             +F^{(\theta)}_{\varphi r}(D_\theta \phi)^{(\theta)}
            \nonumber\\
            & & 
            \left. +F^{(r)}_{\theta \varphi}(D_r \phi)^{(r)}
             +F^{(\theta)}_{\theta \varphi}(D_r \phi)^{(\theta)}
             \right]^2 \frac{1}{\sin^2 \theta r^4} \ , \nonumber\\
%\end{eqnarray}            
 & &\\
%\begin{eqnarray}
{\cal H}_3^{(2)} &=& 6\left\{
\frac{1}{r^2}
\left[
2(\eta^2-\frac{1}{4}(\phi_1^2+\phi_2^2)) F^{(\varphi)}_{r \theta} 
   +(D_r\phi)^{(r)}(D_\theta\phi)^{(\theta)} 
   -(D_r\phi)^{(\theta)}(D_\theta\phi)^{(r)} 
\right]^2 \right. \nonumber\\
& &\left. +
\frac{1}{\sin^2 \theta r^2}
\left[
2(\eta^2-\frac{1}{4}(\phi_1^2+\phi_2^2)) F^{(r)}_{\varphi r} 
   +(D_\varphi\phi)^{(\theta)}(D_r\phi)^{(\varphi)} 
\right]^2 \right. \nonumber\\
& &\left. +
\frac{1}{\sin^2 \theta r^2}
\left[
2(\eta^2-\frac{1}{4}(\phi_1^2+\phi_2^2)) F^{(\theta)}_{\varphi r} 
   -(D_\varphi\phi)^{(r)}(D_r\phi)^{(\varphi)} 
\right]^2 \right. \nonumber\\
& & +\left.
\frac{1}{\sin^2 \theta r^4}
\left[
2(\eta^2-\frac{1}{4}(\phi_1^2+\phi_2^2)) F^{(r)}_{\theta\varphi } 
   +(D_\theta\phi)^{(\theta)}(D_\varphi\phi)^{(\varphi)} 
\right]^2 \right. \nonumber\\
& & +\left.
\frac{1}{\sin^2 \theta r^4}
\left[
2(\eta^2-\frac{1}{4}(\phi_1^2+\phi_2^2)) F^{(\theta)}_{\theta\varphi } 
   -(D_\theta\phi)^{(r)}(D_\varphi\phi)^{(\varphi)} 
\right]^2 \right\} \ ,  \nonumber\\
%\end{eqnarray}            
 & &\\
%\begin{eqnarray}
{\cal H}_2^{(2)} &=& 54(\eta^2-\frac{1}{4}(\phi_1^2+\phi_2^2))^2\left\{
\left[(D_r\phi)^{(r)}\right]^2+\left[(D_r\phi)^{(\theta)}\right]^2
\right. \nonumber\\
& & +\left.
\frac{1}{r^2}\left(
\left[(D_\theta\phi)^{(r)}\right]^2+\left[(D_\theta\phi)^{(\theta)}\right]^2
\right)
+\frac{1}{\sin^2 \theta r^2}\left[(D_\phi\phi)^{(r)}\right]^2
 \right\} \ ,  \nonumber\\
%\end{eqnarray} 
 & &\\
%\begin{eqnarray}
{\cal H}_1^{(2)} &=& 108\left[\eta^2-\frac{1}{4}(\phi_1^2+\phi_2^2)
\right]^4 \ , 
\end{eqnarray} 
respectively. 

With the Ansatz of the gauge fields we express the 
gauge constraint eq.~\re{19} in the form 
\begin{equation}
G_c = \frac{1}{r}(\partial_r H_1 -\frac{1}{r} \partial_\theta H_2)
            = 0 \ .
\end{equation}

The variational equations are to be derived from the Hamiltonians of 
the models with the term $\mu_L G_c^2$ added, where $\mu_L$ is 
the Lagrange multiplier.

For $n=1$ the spherically symmetric Ansatz is recovered 
with 
$H_1(r, \theta) \equiv H_3(r, \theta) \equiv \phi_2(r, \theta) \equiv 0$ and 
$H_2(r, \theta) \equiv H_4(r, \theta) = f(r) \ \ ,
 \ \phi_1(r, \theta)= h(r)$, where $f(r)$, $h(r)$ are
defined in \cite{KOT}.

The boundary conditions follow from the requirements
of finite energy and analyticity as well as symmetry.
They are given by \\
\begin{center}
\begin{tabular}{ccccccccc}
\multicolumn{5}{c}{at the origin} 
&\multicolumn{1}{c}{\hspace*{1.cm}}
&\multicolumn{3}{c}{at infinity}\\
$H_1(0,\theta)$ &=& $H_3(0,\theta)$ &=& 0 & &
$H_k(\infty,\theta)$ &=& 0 \\
$H_2(0,\theta)$ &=& $H_4(0,\theta)$ &=& 1 & &
 \\
$\phi_1(0,\theta)$ &=& $\phi_2(0,\theta)$ &=& 0 & &
$\phi_1(\infty,\theta)/2\eta$ &= &1\\
\end{tabular}
\begin{equation}\label{BC} \end{equation}
%\vspace{1.cm}\\
\begin{tabular}{ccccccccccc}
\multicolumn{5}{c}{on the $z$-axis} 
&\multicolumn{1}{c}{\hspace{3.cm}}
&\multicolumn{3}{c}{on the $\rho$-axis}\\
$H_1(r,0)$ &=& $H_3(r,0)$ &=& 0 & &
$H_1(r,\frac{\pi}{2})$ &=& $H_3(r,\frac{\pi}{2})$ &=& 0 \\
$\partial_\theta H_2(r,0)$ &=& $\partial_\theta H_4(r,0)$ &=& 0 & &
$\partial_\theta H_2(r,\frac{\pi}{2})$ &=& 
$\partial_\theta H_4(r,\frac{\pi}{2})$ &=& 0 \\
$\phi_2(r,0)$ &=& $\partial_\theta\phi_1(r,0)$ &=& 0 & & 
$\phi_2(r,\frac{\pi}{2})$ &=& 
$\partial_\theta\phi_1(r,\frac{\pi}{2})$ &=& 0 \\
\end{tabular}
\end{center}

\section{Numerical results}

This Section deals with the results of the numerical integrations. The
topological charge-1 and charge-2 solutions of the systems under consideration
will be found numerically and the energies {\it per unit topological charge}
 of these
solitons will be calculated as functions of the dimensionless coupling
constants $\lambda_A$ with the aim of deducing the presence of attractive and
repulsive phases as the case may be, following the procedure of Ref.~\cite{JR}.
This will be done by holding all but one of the dimensionless coupling
constants constant and varying the last one only, although, given that
our models are charcterised with more than one such constant, it would be
possible to vary more than one of these at a time. This we do not do here.
The present Section consists of three subsections.

The first subsection is concerned with certain extended
$p=1$ models. Of these the first is the GG model, while the next five are
perturbative extensions of the $p=1$ model which were described in Section
{\bf 2.3.2} as perturbed $p=1$ composite models consisting of the $p=1$
model and all or some of the terms in the $p=2$ model. 
The latter therefore differ from the
GG model in two respects. Firstly, they incorporate terms from the $p=2$ model
which the GG model does not, and secondly, these terms may  be regarded as
perturbations of the $p=1$ sytem unlike the GG model as explained in Section
{\bf 2.3.2} and in Ref.~\cite{KKT}. In particular, we have studied the $p=1$
model extended by each of the four terms constituting the $p=2$ model, 
separately.
The interest in these (perturbative) models is that according to 
our analysis in
section {\bf 2.3.2} the attraction or repulsion of two like monopoles is
decided by the slope of the energy versus the $\lambda_A$ in question, and
assuming quite reasonably that these curves are monotonic, 
this yields systems
supporting only an attractive phase, or only a repulsive one. 
In addition to these
four examples, we have studied the case of two nonvanishing $\lambda_A$, 
each
chosen such that in the previous analysis the model 
with that particular $\lambda_A$
supported, respectively, an attractive and a repulsive phase. 
In such a case, it is intuitively
expected that the combined terms will result in a model 
supporting both phases,
and since both $\lambda_A$ are very small, the crossover point of the two
phases is expected to be very near $\lambda_A =0$, the critical value of the
model, which in this case coincides with the critical value of the PS model. 

The second subsection deals with the generic case of the composite model,
i.e. that in which the values of the coupling constants $\lambda_A$ are not
small, so that this is not a perturbative extension of the $p=1$ model.
We will set three of the four dimensionless coupling constants $\lambda_A$
equal to 1 and varying the fourth one. 
The main interest in these
models is the question as to what phases are supported, the models themselves
not being as closely {\it almost self-dual} as the $p=2$ model on its own. 
For
that reason it is not so easy to predict where, if anywhere, 
the crossover points
will be situated. The other, technical, interest in the full composite model is
that it affords us with an approach to the numerical integration of the $p=2$
model on its own, by gradualy decreasing its $p=1$ content. 

The third subsection deals exclusively with the $p=2$ model, which
from the viewpoint of numerical integrations is the most
difficult of the models considered here.

The partial differential equations are solved numerically 
using the iterative Newton-Raphson method \cite{cadsol}.
The calculations are performed with the rescaled Higgs field functions
$\phi_1(r,\theta)/\eta$, $\phi_2(r,\theta)/\eta$.
We employed
the compactified dimensionless radial coordinate 
$x=\eta r/(1+\eta r)$ to map spatial infinity to the finite
value $x=1$. 
The differential equations are discretized on a non-equidistant 
grid in $x$ and $\theta$ covering the integration region 
$0\le x \le 1$, $0 \le \theta \le \pi/2$. Typical grids have sizes 
of $70 \times 20$ gridpoints. The error estimates are of 
the order $10^{-3}$.

\subsection{Extended $p=1$ models}

\subsubsection{The Georgi-Glashow model}

This is the best known extended $p=1$ model, and has been extensively studied
in the literature analytically. A thorough numerical study of this model was
recently performed in \cite{KKT} where it was confirmed that this system
supports a repulsive phase only. We refer to \cite{KKT} for the analysis of
this model.

This system consists of the $p=1$ model to
which the usual symmetry breaking Higgs self-interaction potential is added.
It is well known~\cite{N,KKT} that the addition of the latter term cannot be
regarded as a perturbation of the $p=1$ model. In this respect this model
differs from the composite models following in the next sections.

In Fig.~1 we show the energy density of the charge-2
solution of the $p=1$ model as a function 
of the coordinates $\rho = \eta r \sin \theta$ and $z= \eta r \cos \theta$.
The energy density has its maximum on the $\rho$-axis and is monotonic
decreasing on the $z$-axis. Thus its shape deviates considerably form 
spherical symmetry.
The surfaces of equal energy density form spheres for large distances form
the origin and doughnuts for distances more close to the origin.
The Maximum of the energy density corresonds is located on a 
circle in the $x-y$ plane.

\subsubsection{$\lambda_4 =\lambda_3 =\lambda_2 =0$ perturbed composite model}

In this model only the potential term
$\frac{5}{18\times 32}\eta^{-4}\lambda_1 {\cal H}_1^{(2)}$ 
is added to the $p=1$ model.
In Fig.~2 the quantity $\Delta E(\lambda_1)$, as defined in eq.~(\ref{16}), 
is plotted 
as a function of the coupling constant 
$\lambda_1$ (solid line {\bf a}) for $0\leq \lambda_1 \leq 1$.
$\Delta E(\lambda_1)$ is a monotonically increasing function.
Consequently there is only a repulsive phase in this model.

The dashed straight line in Fig.~2 corresponds to the linear 
approximation of $\Delta E$ defined in eq.~(\ref{18}).
The deviation of the function $\Delta E(\lambda_1)$
from this line occurs at rather small values of $\lambda_1$. 
This shows that  the linear approximation  is valid only 
for very small values of $\lambda_1$, roughly in the range 
$0 \le \lambda_1 < 1/100$.

Note, that for this model the analysis of ref.~\cite{KS} can be adopted.
The reason is, that the actual form of 
the Higgs potential does not enter the analysis of the zero modes.
Thus the conclusions of ref.~\cite{KS} are valid for this model, too,
provided the Higgs potential forces the  Higgs field to decay faster 
than $1/r$.
 
\subsubsection{$\lambda_4 =\lambda_3 =\lambda_1 =0$ perturbed composite model}

This model consists of the $p=1$ model with only the term
$\frac{5}{18\times 32}\eta^{-4}\lambda_2{\cal H}_2^{(2)}$ 
added to 
the $p=1$ model.
The quantity $\Delta E(\lambda_2)$ is shown in Fig.~2 as a function of 
$\lambda_2$
(solid line {\bf b}). Again, $\Delta E(\lambda_2)$ is a
monotonically increasing function.
Thus there is only a repulsive phase in this model.

The dashed straight line shows the linear approximation of $\Delta E$, 
eq.~(\ref{18}). 
For $\lambda_2>0.1$, $\Delta E$ deviates from this line.
Therefore, the validity of the linear approximation to this model
is restricted to the range $0\le \lambda_2 < 0.1$ 

\subsubsection{$\lambda_4 =\lambda_2 =\lambda_1 =0$ perturbed composite model}

This model consists of the $p=1$ model with only the term 
$\frac{5}{18\times 32}\eta^{-4}\lambda_3{\cal H}_3^{(2)}$
added to the $p=1$ model. This Skyrme like term leads to 
an attractive phase as can be seen form Fig.~2 (solid line {\bf c}).
$\Delta E(\lambda_3)$ is a monotonically decreasing function.
The deviation from the linear approximation
(dashed line) occurs at $\lambda_3=0.1$ implying that
the validity of the linear approximation to this model
is restricted to the range $0\le \lambda_3 < 0.1$.

\subsubsection{$\lambda_3 =\lambda_2 =\lambda_1 =0$ perturbed composite model}

This model consists of the $p=1$ model with only the term 
$\frac{5}{18\times 32}\eta^{-4}\lambda_4{\cal H}_4^{(2)}$ added.
Again this model has only an attractive phase. The quantity
$\Delta E(\lambda_4)$ (solid line {\bf d}) as a function of
$\lambda_4$ is shown in Fig.~2 together with the
linear approximation (dashed line).
The linear approximation is valid for $0 \le \lambda_4 < 0.2$.

\subsubsection{$\lambda_4 =\lambda_2 =0$  composite model}

The numerical results for the composite models with only one of 
the terms eq.~(\ref{2.1}-\ref{2.4}) added to the $p=1$ model 
show that there exist
models with attractive or repulsive phases only. 
On the other hand, from the analysis of section {\bf 2.3.2} 
follows that for small coupling constants $\lambda_A$ 
models with both attractive and repulsive phases 
can exist, provided at least one of the terms added to the $p=1$ model
leads to a repulsive phase and the other one
leads to an attractive phase. 

We studied numerically a model where the terms 
$\frac{5}{18\times 32}\eta^{-4}\lambda_1 {\cal H}_1^{(2)}$ 
and 
$\frac{5}{18\times 32}\eta^{-4}\lambda_3{\cal H}_3^{(2)}$ 
are added to the $p=1$ model.
In Fig.~3a we show the energy {\it per unit topological charge} 
$E(\lambda_1, \lambda_3)/n$
for the charge-1 and charge-2 solutions as functions of $\lambda_3$ with
$\lambda_1=0.01$ fixed.
Both functions increase with $\lambda_3$. For small values of $\lambda_3$
the energy {\it per unit topological charge}
of the charge-2 solution (dashed-dotted line) is 
larger then the energy {\it per unit topological charge}
of the charge-1 solution (solid line), whereas for larger values of
$\lambda_3$ the energy {\it per unit topological charge} of the 
charge-1 solution is larger than the 
energy {\it per unit topological charge} of the charge-2 solution.
The crossover occurs at $\lambda_3 \sim 0.22$ where both solutions 
have the same energy {\it per unit topological charge} 
$E/n \sim 1.085 \times 4 \pi \eta$. Notice that attraction/repulsion
occur for values of $\lambda_3$ above/below the crossover point, in
contrast with the situation~\cite{JR} in the Abelian Higgs model.

As expected, this model supports an attractive and an repulsive phase.
The crossover is at a small value of the coupling constants $\lambda_A$
and the energy {\it per unit topological charge} at this point is
near the Bogolmol'nyi bound.
However, in the example chosen the crossover is outside the 
range where the linear approximation is valid. From the linear approximation
we can only predict the existence of attractive/repulsive phases, but 
not the features of the models in detail.

Another example is shown in Fig.~3b. Here 
the energy {\it per unit topological charge} is shown for 
the charge-1 solution (solid line) and the charge-2 solution 
(dashed-dotted line) as a function of $\lambda_1$ for fixed
coupling constant $\lambda_3=1$. 
Again, for both solutions the energy {\it per unit topological charge}
is an increasing function. There is an attractive phase for
small values of $\lambda_1$ and a repulsive phase for large values of 
$\lambda_1$. The crossover is at $\lambda_1 \sim 0.07$ where the 
energy {\it per unit topological charge} of both solutions 
is $E/n=1.375\times 4 \pi \eta$. 
For large values of the coupling constant, like $\lambda_3=1$, the 
linear approximation is certainly not valid anymore. Still the
model supports both an attractive and a repulsive phase. However, 
the energy {\it per unit topological charge} at the crossover now
exceeds the Bogolmol'nyi  bound considerably.

\subsection{Composite model}

In this section we consider the model defined by the Hamiltonian
eq.~(\ref{10}), where the parts of the $p=1$ model and the $p=2$ 
model are treated on the same footing. The linear approximation 
of the energy {\it per unit topological charge} as a function 
of the coupling parameters $\lambda_A$ is not applicable for this 
model. 

As a typical example we discuss the charge-2 solution for the 
coupling constants $\lambda_A=1$, $A=1,\dots ,4$.
The energies {\it per unit topological charge}
of the charge-2 and charge-1 solutions are $E_{n=2}/2=2.1433 \times 4 \pi \eta$ 
and $E_{n=1}=2.05993 \times 4 \pi\eta$, respectively.
Thus the charge-2 solution is 
in the repulsive phase for these values of the coupling constants.
 
In Figs.~4a-4g we show the energy density and the 
field configuration for the charge-2 solution as functions of the 
dimensionless coordinates $\rho$ and $z$.
In Fig.~4a the energy density is shown. It possesses
a maximum on the $\rho$-axis and it is monotonically decreasing
along the $z$-axis, indicating a strong deviation form spherical symmetry.
Like for the $p=1$ model 
the surfaces of equal energy density form spheres und doughnuts.

Let us now compare the energy densities 
of the charge-2 solutions in
the composite model and in the $p=1$ model, shown in Fig.~1.
At the origin
the energy density of the composite model is roughly 
one order of magnitude larger than the energy density of the $p=1$ model. 
In the composite model the energy density approaches its
asymptotic value at a shorter distance form the origin than
in the $p=1$ model.
Thus the composite model possesses a core with a high energy density, 
whereas the core in the $p=1$ model is softer and more extended.

In Fig.~4b the gauge field function $H_1(\rho,z)$ is shown. According to
the boundary conditions $H_1(\rho,z)$ vanishes on the $\rho$ and $z$ axis.
It has a maximum located near the origin. For spherically symmetric
solutions $H_1(\rho,z)$ would vanish identically. Thus $H_1(\rho,z)$ is 
excited due to the axial symmetry of the solution.
Fig.~4c shows the gauge field function $H_2(\rho,z)$. The deviation from
spherical symmetry for this function is small near the origin and
nearly vanishing far away from the origin.
The function $H_3(\rho,z)$ is exhibited in Fig.~4d. Simmilarly to the 
function $H_1(\rho,z)$ shown in Fig.~4b this function is excited due to 
the axial symmetry of the solution. Its maximum is located near the origin.
Note that it is roughly one order of magnitude smaller than the maximum of
the function $H_1(\rho,z)$. 
In Fig.~4e the function $H_4(\rho,z)$ is shown. Like the function 
$H_2(\rho,z)$ shown in Fig.~4.c the deviation from sphercial
symmetry is only small for this function. 
The Higgs field function $\phi_1(\rho,z)$ is shown in Fig.~4f.
At the origin this function is linear in  $z$ for $\rho=0$ whereas 
for $z=0$ it is quadratic in $\rho$. For large distances from the
origin the deviation from spherical symmetry almost vanishes.
Fig.~4g shows the Higgs field function $\phi_2(\rho,z)$. It has a minimum
near the origin and vanishes on the $\rho$ and $z$ axis due to the 
boundary conditions. Like the gauge field functions $H_1(\rho,z)$ and  
$H_3(\rho,z)$ this function is excited by the axial symmetry of the solution.

The gauge field functions and Higgs field functions possess a non-trivial
dependence on the coordinates $\rho$ and $z$ only in the region 
near the origin where the core of the energy density is located.
As a result of our numerical integrations we have found that
outside this core the functions decay to their 
vacuum values at infinity at a rate faster than a power behaviour. 
This is in contrast to the solutions of the $p=1$ model,
where in absense of the Higgs potential the Higgs fields possess
a power law decay and 
only the gauge field functions decay exponentially. 
Thus for the composite model the asymptotic decay of the energy density 
outside the core originates not from the Higgs field functions
but from the gauge field functions only.

Let us next look at the dependence of the 
energy on the coupling constants
$\lambda_A$ for  the charge-1 and charge-2 solutions of the composite model.

We have solved numerically the partial differntial equations for fixed
$\lambda_4=1$ and several values of $\lambda_A$ in the range 
$0 \leq \lambda_A\leq 1.5$
for $A=1,2,3$.
The results are exhibited in Fig.~5.
Lines {\bf a} and {\bf a'} represent the energy {\it per unit topological
charge} of the charge-1 and charge-2 solutions, respectively, as  
functions of the coupling constant $\lambda_1$ with
$\lambda_2 =\lambda_3 =1$ fixed. The lines cross at $\lambda_1 \sim 0.1$.
Thus there is both an attractive phase ($\lambda_1 < 0.1$)
and a repulsive phase ($\lambda_1 > 0.1$). 
The value of the energy {\it per unit topological charge} 
$E/n=1.865\times 4\pi \eta$
at the crossover is 
rather large compared to the Bolgolmol'nyi bound, which 
takes the value $1.1 \times 4\pi\eta$ for $\lambda_1=0.1$ in this case.

Lines {\bf b} and {\bf b'} represent the 
energy {\it per unit topological charge} 
of the charge-1 and charge-2 solutions, respectivley, 
when $\lambda_2$ is varied with fixed $\lambda_1=\lambda_3=1$.
In the range $0 \leq \lambda_2 \leq 1.5$ the lines {\bf b} and {\bf b'}
are almost parallel and there is no crossover. 
The upper lines correspond to the charge-2 solutions. There is
a repulsive phase only in this case. 

Finally we fix $\lambda_1=\lambda_2=1$ and vary $\lambda_3$. 
The energy {\it per unit topological charge}
is represented in Fig.~5 by lines {\bf c} and {\bf c'} for the 
charge-1 and charge-2 solutions, respectively.
Again, lines {\bf c} and {\bf c'} are almost parallel. 
Thus, there is only a repulsive phase.

\subsection{$p=2$ model}

The numerical integration of the variational equations of the 
$p=2$ model is a very difficult task. 
In the absense of the terms of the $p=1$ model numerical
instabilities arise leading to a failure of the 
algorithms.
We succeded to construct the spherically symmetric charge-1 solutions.
However, we did not find the axially symmetric charge-2 solutions 
of the $p=2$ model on its own.
Instead we used an approximation procedure to calculate the 
energy of the charge-2 solutions as explained in the following.

We start with the composite model and regard now the 
$p=1$ part as a perturbation. For that purpose we modify the
static Hamiltonian eq.~(\ref{10}) by introducing the new coupling
constant $\lambda_0$,
\be
\label{10a}
\hat {\cal {H}}^{(c)}=\lambda_0 {\cal H}^{(1)} +\eta^{-4}{\cal H}^{(2)},
\ee
and consider the limit $\lambda_0 \rightarrow 0$. Assuming the existence
of the solutions to the $p=2$ model, we will now consider the model
\re{10a} as a perturbation of the latter by the $p=1$ model,
characterised by a small value of the dimensionaless coupling
constant $\lambda_0 \ll 1$.

We expand the energy of the magnetic charge-$n$ soliton in 
the vicinity of the parameter
$\lambda_0$,
\be
\label{12x}
E^{(n)}(\lambda_0+\Delta \lambda_0) = E^{(n)}(\lambda_0)
+\Delta\lambda_0 \left.
\frac{dE^{(n)}}{d\lambda_0}\right|_{\lambda_0}
+o(\lambda_0^2) .
\ee
With the same reasoning as in section {\bf 2.3.2},
the quantity $\left.\frac{dE^{(n)}}{d\lambda_0}\right|_{\lambda_0 =0}$ can be
calculated exclusively in terms of the exact solutions.
Now, because we assumed that $\lambda_0$ is very small, we can evaluate
(\ref{12x}) at $\Delta \lambda_0 = -\lambda_0$. 
Thus for sufficiently small $\lambda_0$,
\begin{equation}
E^{(n)}(0) \approx E^{(n)}(\lambda_0) - \lambda_0
\left. \frac{dE^{(n)}}{d\lambda_0}\right|_{\lambda_0}
\end{equation}
will give a very good approximation to the energy of a solution 
of the $p=2$ model.
This energy consists of two terms, the first  is
the energy of the modified $p=2$ model for a small value of
$\lambda_0$. 
The second term is given by $-\lambda_0 \int {\cal H}^{(1)} d^3r$ 
evaluated for the solution with finite (but small) $\lambda_0$.

We can now estimate the energy of the $p=2$ model on its own 
by calculating the energy of the modified composite model for
a very small value of $\lambda_0$ and by subtracting the contribution 
of the $p=1$ part, i.~e. the quantity $\lambda_0 \int{\cal H}^{(1)}d^3r$ 
evaluated with the solution for the small value of $\lambda_0$.

To check the validity of our limiting scheme,
we followed this procedure for a set of decreasing values of 
$\lambda_0$ and found, that the estimated energy becomes indeed a constant
for $\lambda_0$ small enough.

Let us first we discuss the properties of charge-2 solutions 
for the $p=2$ model at the {\it critical} values of the 
coupling constants  $\lambda_A=1$, $A=1, \dots ,4$, determined by the 
investigation of the charge-1 solutions.
The energy {\it per unit topological
charge} of the charge-1 solution was found to be 
$E^{(1)}(\lambda_A=1)=1.0002 \times 4 \pi \eta$, whereas 
for the charge-2 solutions we find for the same coupling constants 
$E^{(2)}(\lambda_A=1)/2=1.003 \times 4 \pi \eta$.
Thus the like monopoles repel for the {\it critical} values of the
coupling constants. 
If we compare with the Bogolmol'nyi bound, given by $4 \pi \eta$,
we see that for both the charge-1 and charge-2 solutions the energies 
{\it per unit topological charge}
are very close to the lower bound, indeed.

In order to compare the $p=2$ model with
the $p=1$ model and the composite model we
show in Fig.~6a-6g the energy density,  
the gauge field functions and the Higgs field functions for the 
solution of the modified model (\ref{10a}) for the coupling constants
$\lambda_0=0.01$, $\lambda_A=1$, $A=1, \dots ,4$ and assume that this
configuration is close to the solution of the $p=2$ model on its own.

The energy density, shown in Fig.~7a as a function of the dimensionless
coordinates $\rho$ and $z$, is similar to the energy density of the composite
model, shown in Fig.~4a. It possesses a maximum on the $\rho$-axis 
near the origin and it is monotonically decreasing along the $z$-axis. 
Its value at the origin is of the same magnitude as for the 
composite model and the decay for large distances from the 
origin is slightly faster. 
Like for the $p=1$ model and the composite model
the surfaces of equal energy density form spheres und doughnuts.

The gauge field functions and the Higgs field functions are shown in 
Figs.~6b-6e and Figs.~6f-6g, respectively.
They look very simmilar to the solution of the composite model
shown in Figs.~4b-g, except for the gauge field function $H_3(\rho,z)$.
For the $p=2$ model  $H_3(\rho,z)$  is negative and its minimum is of 
the order $\approx -0.1$, whereas for the composite model it is positive with 
a maximum of the order $\approx 0.025$.

Let us next look at the dependence of the energy on the 
coupling constants $\lambda_A$ for the charge-1 and charge-2 solutions.
For the $p=2$ model we have fixed the coupling constant $\lambda_4=1$.
For the charge-1 solutions we solved the ordinary differential equations
of the $p=2$ model for several values of the coupling constants. 
The energy for the charge-2 solutions are calculated approximately as 
discussed above.

The results are shown in Fig.~7a-7c.
In Fig.~7a we exhibit the energy {\it per topological charge} of 
the charge-1 solution (solid line) and the charge-2 solution (dashed-dotted
line) as a function of the coupling constant $\lambda_1$ with fixed 
$\lambda_2=\lambda_3=1$.
Both lines cross at $\lambda_1=0.89$. 
There is an attractive phase for $\lambda_1<0.89$ and a repulsive phase
for $\lambda_1>0.89$. The energy {\it per topological charge} at the 
crossing is $E/n=0.9898 \times 4 \pi \eta$. 
As expected the crossover takes place 
at values of the coupling constants near the {\it critical} values
$\lambda_1=\lambda_2=\lambda_3=1$.

In Fig.~7b we exhibit the energy {\it per topological charge} of 
the charge-1 solution (solid line) and the charge-2 solution (dashed-dotted
line) as a function of the coupling constant $\lambda_2$ with fixed 
$\lambda_1=\lambda_3=1$.
In this case the crossing is at  $\lambda_2=1.07$, with a repulsive phase
for $\lambda_2< 1.07$ and an attractive phase for $\lambda_2> 1.07$. For the
energy {\it per topological charge} at the crossing
the value $E/n=1.0333 \times 4 \pi \eta$ was found.
Again, the values of the coupling constants at the crossing are
near the {\it critial} values.

In Fig.~7c we exhibit the energy {\it per topological charge} of 
the charge-1 solution (solid line) and the charge-2 solution (dashed-dotted
line) as a function of the coupling constant $\lambda_3$ with fixed 
$\lambda_1=\lambda_2=1$.
In this case the crossing occurs for a value of the
coupling constants $\lambda_3=1.49$, which is 
considerably larger then the {\it critical} value $\lambda_3=1$.
There is a repulsive phase for $\lambda_3<1.49$ and an attractiv phase
for $\lambda_3>1.49$. The energy {\it per topological charge} at the 
crossing takes the value $E/n=1.1969 \times 4\ pi \eta$.

Notice that the situation in Figs.~7b and 7c, namely that of
attraction/repulsion taking place for values of the coupling constants
larger/smaller than that of the crossover point, contrasts with
the corresponding situation~\cite{JR} for the Abelian Higgs model.

\section{Summary and Conclusions}

The main aim of the present paper was to acheive monopole field
configurations in which like-monopoles can attract as well as repel.
We found this to be the case in various new 3 dimensional $SO(3)$ gauged
Higgs models with isovector Higgs fields.
Such a property contrasts with the 't~Hooft-Polyakov monopoles of the
Georgi-Glashow model, in which like-monopoles only repel.
The main model we employed was one we described
as the $p=2$ model because it is the second one in a hierarchy of models
descended from $4p$ dimensional YM systems~\cite{T},
the first one of which, $p=1$, is the GG model in the PS limit. The
spherically symmetric magnetic charge-1 solutions of the $p=2$ model were
previously studied, from which we concluded that the type of behaviour
sought in the present work was likely to materialise. Accordingly in
the present investigation, we studied the charge-2 axially symmetric
solutions of this model, and found indeed that for some values of the
dimensionless coupling parameters the interaction energies of the
monopoles led to bound states. In this sense, the 3 dimensional $p=2$
model is very similar to the 2 dimensional Abelian Higgs model, both
featuring a crossover between an attractive and a repulsive phase.
Indeed, our procedures were the same employed by Jacobs and
Rebbi~\cite{JR} in the 2 dimensional case. There is however, a qualitative
difference between these models in the sense that the attractive and
repulsive phases of the $p=2$ model occur on the opposite sides of the
crossover points for some of the coupling constants.
Another difference between these models is that in the $p=2$ case there is
really no genuinely non-interacting, minimal, field configuration.
Nevertheless there is an almost non-interacting field configuration
which was discussed in some detail in \cite{KOT}, which
turns out to be approximately the crossover point between the two phases. 

In addition to the $p=2$ model, we studied combinations of the $p=1$,
and the $p=2$ models, as well as extensions of the $p=1$ model by some
of the constituent terms in the $p=2$ model
which is tantamount to adding a particular Skyrme-like
term. In some of the latter
examples we also considered small perturbations of the $p=1$ model,
something which is radically different from adding the Higgs potential
to the $p=1$ model to obtain the GG model. ( Recall that the BPS solutions
cannot be related to the monopoles of the GG model perturbatively.)
Perhaps the most interesting outcome of the latter considerations was
the construction of extended $p=1$ models whose monopole solutions
feature either an attractive phase only or a repulsive phase only,
between two like-monopoles.

Apart from their intrinsic interest in providing generalised monopoles
exhibiting both attractive and repulsive phases, it is hoped that this
property of the generalised monopoles may be of some potential interest
in the role of monopoles in grand unified theory, and possibly
to the cosmic monopole problem~\cite{K1}. The models we have considered
could occur as extensions of the usual models of the grand unified
theory. In such models there is  attraction between like-monopoles
(and ofcourse like-antimonopoles) as happens with
type I cosmic strings~\cite{K2}. We do not speculate here on the possible
consequences of this analogy. Perhaps the most likely effect of such
(generalised) monopoles may occur in the gravitational clumping~\cite{GKT}
of monopoles leading to the annihilation of monopole-antimonopole pairs.
In contrast with the usual theory, there will be only attractive forces
between the monopoles, independently of the gravitational interaction.

\bigskip

{\bf Acknowledgements}
This work was supported by
Basic Science SC/96/636 and SC/95/602 grants of Forbairt.

\newpage

\noindent
\newpage

\small{

 }
\newpage
\begin{center}
{\bf Figure captions }
\end{center}
{\bf Fig.~1}\\
The energy density in units of $\eta^4$ of the $p=1$ model 
is shown as a function
of the dimensionless coordinates 
$\rho=\eta r \sin \theta$ and $z=\eta r \cos \theta$.
\vspace{5.mm}\\
{\bf Fig.~2}\\
The quantity $\Delta E(\lambda_A)=E^{(2)}(\lambda_A)/2-E^{(1)}(\lambda_A)$
in units of $4\pi \eta$
(solid lines) is shown as a function of $\lambda_A$ with 
$\lambda_B=0$, $B\neq A$. The dash-dotted lines indicate the 
slope at $\lambda_A=0$. 
Line {\bf a} corresponds to $\lambda_1 \neq 0$,
line {\bf b} to $\lambda_2 \neq 0$, 
line {\bf c} to $\lambda_3 \neq 0$ and 
line {\bf d} to
$\lambda_4 \neq 0$, respectively.
\vspace{5.mm}\\
{\bf Fig.~3a}\\
The energy {\it per unit topological charge} 
in units of $4\pi \eta$
is shown as a function of $\lambda_3$
with $\lambda_1=0.01$, $\lambda_2=\lambda_4=0$ for the charge $n=1$ solution 
(solid line) and for the charge $n=2$ solution (dash-dotted line).
\vspace{5.mm}\\
{\bf Fig.~3b}\\
The same as Fig.~3a for $\lambda_1$ with $\lambda_3=1$ and
$\lambda_2=\lambda_4=0$.
\vspace{5.mm}\\
{\bf Fig.~4a}\\
The energy density in units of $\eta^4$
of the charge $n=2$ solution of the
composite model with $\lambda_A=1$, $A=1,\dots,4$
is shown as a function of the dimensionless coordinates 
$\rho=\eta r \sin \theta$ and $z=\eta r \cos\theta$.
\vspace{5.mm}\\
{\bf Fig.~4b}\\
The same as Fig.~4a for the gauge field function $H_1(\rho,z)$.
\vspace{5.mm}\\
{\bf Fig.~4c}\\
The same as Fig.~4a for the gauge field function $H_2(\rho,z)$.
\vspace{5.mm}\\
{\bf Fig.~4d}\\
The same as Fig.~4a for the gauge field function $H_3(\rho,z)$.
\vspace{5.mm}\\
{\bf Fig.~4e}\\
The same as Fig.~4a for the gauge field function $H_4(\rho,z)$.
\vspace{5.mm}\\
{\bf Fig.~4f}\\
The same as Fig.~4a for the Higgs field function $\phi_1(\rho,z)$.
\vspace{5.mm}\\
{\bf Fig.~4g}\\
The same as Fig.~4a for the Higgs field function $\phi_2(\rho,z)$.
\vspace{5.mm}\\
{\bf Fig.~5}\\
The energy {\it per unit topological charge} 
in units of $4 \pi \eta$
of the composite model 
is shown as a function of $\lambda_A$, $A=1,2,3$ with fixed
$\lambda_B=1$, $B\neq A$ for the charge
$n=1$ solutions (solid lines) and 
for the charge $n=2$ solutions (dash-dotted lines).
Lines {\bf a} and {\bf a'} correspond to 
$\lambda_2=\lambda_3=\lambda_4 = 1$,
lines {\bf b} and {\bf b'}  to 
$\lambda_1=\lambda_3=\lambda_4 = 1$, 
and lines {\bf c} and {\bf c'}  to 
$\lambda_1=\lambda_2=\lambda_4 = 1$,
respectively. 
The thin solid line indicates the Bogolmol'nyi bound.
\vspace{5.mm}\\
{\bf Fig.~6a}\\
The energy density in units of $\eta^4$
of the charge $n=2$ solution of the
$p=2$ model with $\lambda_A=1$, $A=1,\dots,4$
is shown as a function of the dimensionless coordinates 
$\rho=\eta r \sin \theta$ and $z=\eta r \cos \theta$.
\vspace{5.mm}\\
{\bf Fig.~6b}\\
The same as Fig.~6a for the gauge field function $H_1(\rho,z)$.
\vspace{5.mm}\\
{\bf Fig.~6c}\\
The same as Fig.~6a for the gauge field function $H_2(\rho,z)$.
\vspace{5.mm}\\
{\bf Fig.~6d}\\
The same as Fig.~6a for the gauge field function $H_3(\rho,z)$.
\vspace{5.mm}\\
{\bf Fig.~6e}\\
The same as Fig.~6a for the gauge field function $H_4(\rho,z)$.
\vspace{5.mm}\\
{\bf Fig.~6f}\\
The same as Fig.~6a for the Higgs field function $\phi_1(\rho,z)$.
\vspace{5.mm}\\
{\bf Fig.~6g}\\
The same as Fig.~6a for the Higgs field function $\phi_2(\rho,z)$.
\vspace{5.mm}\\
{\bf Fig.~7a}\\
The energy {\it per unit topological charge} in units of $4 \pi \eta$
of the $p=2$ model 
is shown as a function of $\lambda_1$
with fixed $\lambda_A=1$, $A \neq 1$ for the charge $n=1$ solution 
(solid line) and for the charge $n=2$ solution (dash-dotted Line).
The thin solid line indicate the topological bound.
\vspace{5.mm}\\
{\bf Fig.~7b}\\
The same as Fig.~7a for $\lambda_2$ with fixed $\lambda_A=1$, $A \neq 2$.
\vspace{5.mm}\\
{\bf Fig.~7c}\\
The same as Fig.~7a for $\lambda_3$ with fixed $\lambda_A=1$, $A \neq 3$.
\newpage

\begin{table}[t!]
%$\begin{table}[h!]
\begin{center}
\begin{tabular}{|c|cc|c|} 
 \hline 
\multicolumn{1}{|c|} { $ $ }&
\multicolumn{2}{ c|}  
{$1/n\ dE/d\lambda_A |_{\lambda_A=0}$ } &
%{${\displaystyle\left. \frac{dE}{d\lambda_A}\right|_{\lambda_A=0}}$ } &
\multicolumn{1}{|c|} { $c_A$ }\\
 \hline 
$A$ &  $n=1$  & $n=2$ & $ $  \\
 \hline 
$1$ &  $0.688$   & $2.925$  & $2.237$    \\ 
$2$ &  $0.383$   & $0.437$  & $0.054$    \\ 
$3$ &  $0.367$   & $0.291$  & $-0.076$    \\  
$4$ &  $0.061$   & $0.020$  & $-0.041$    \\ 
\hline  
\end{tabular}
\end{center} 
{\bf Table 1}\\
The quantity 
$\left.\frac{1}{n} \frac{dE^{(n)}}{d\lambda_A}\right|_{\lambda_A =0}$
defined in (\ref{15}) for the Hamitonians ${\cal H}_A^{(2)}$
is given for the charge-1 and charge-2 solutions
together with their difference $c_A$ in units of $4 \pi \eta$.
\end{table}
\clearpage
\newpage

\begin{figure}
\centering
\vspace{-1cm}
\mbox{  \epsfysize=12.5cm \epsffile{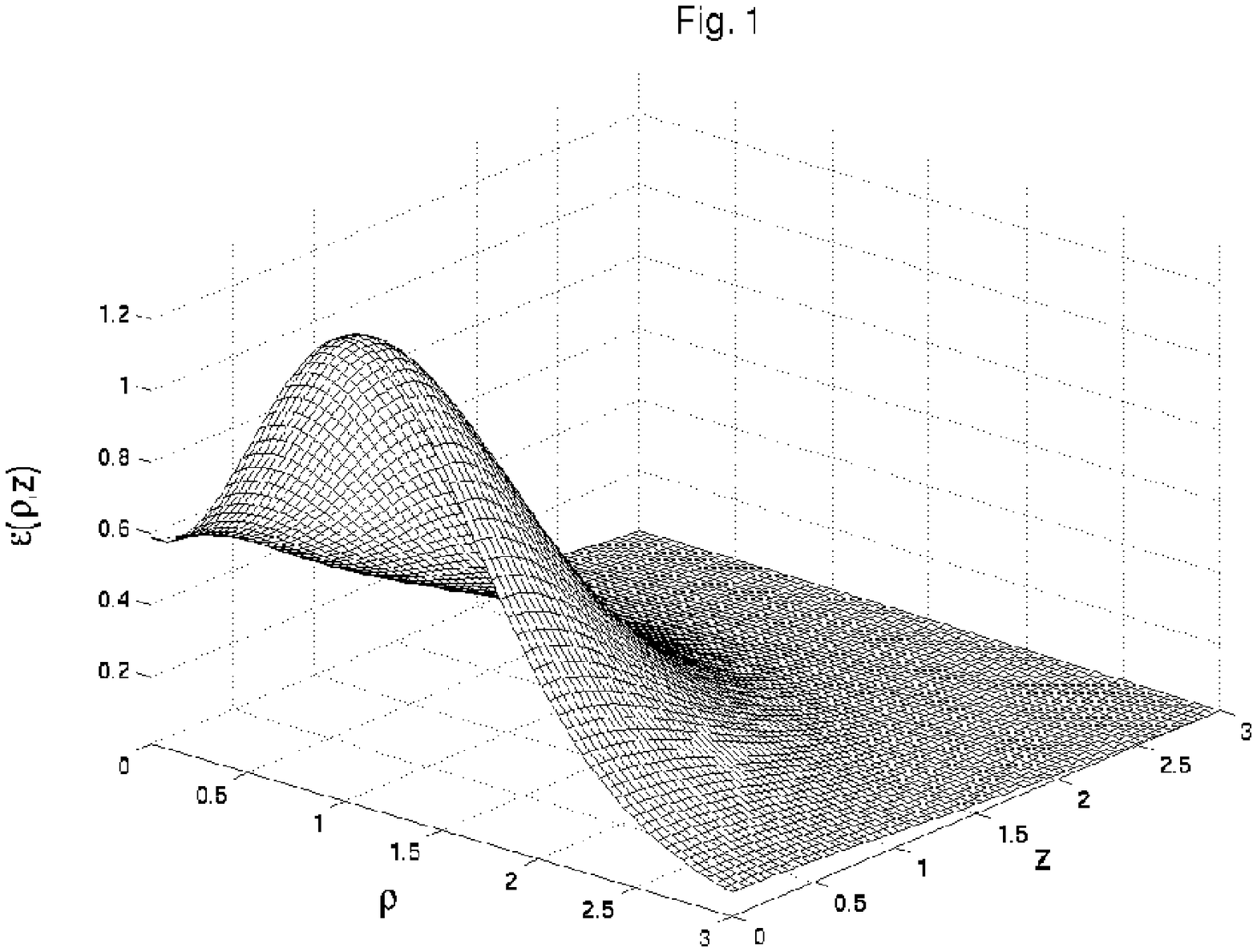}}\\
\end{figure}
\clearpage
\newpage

\begin{figure}
\centering
\vspace{-1cm}
\mbox{  \epsfysize=12.5cm \epsffile{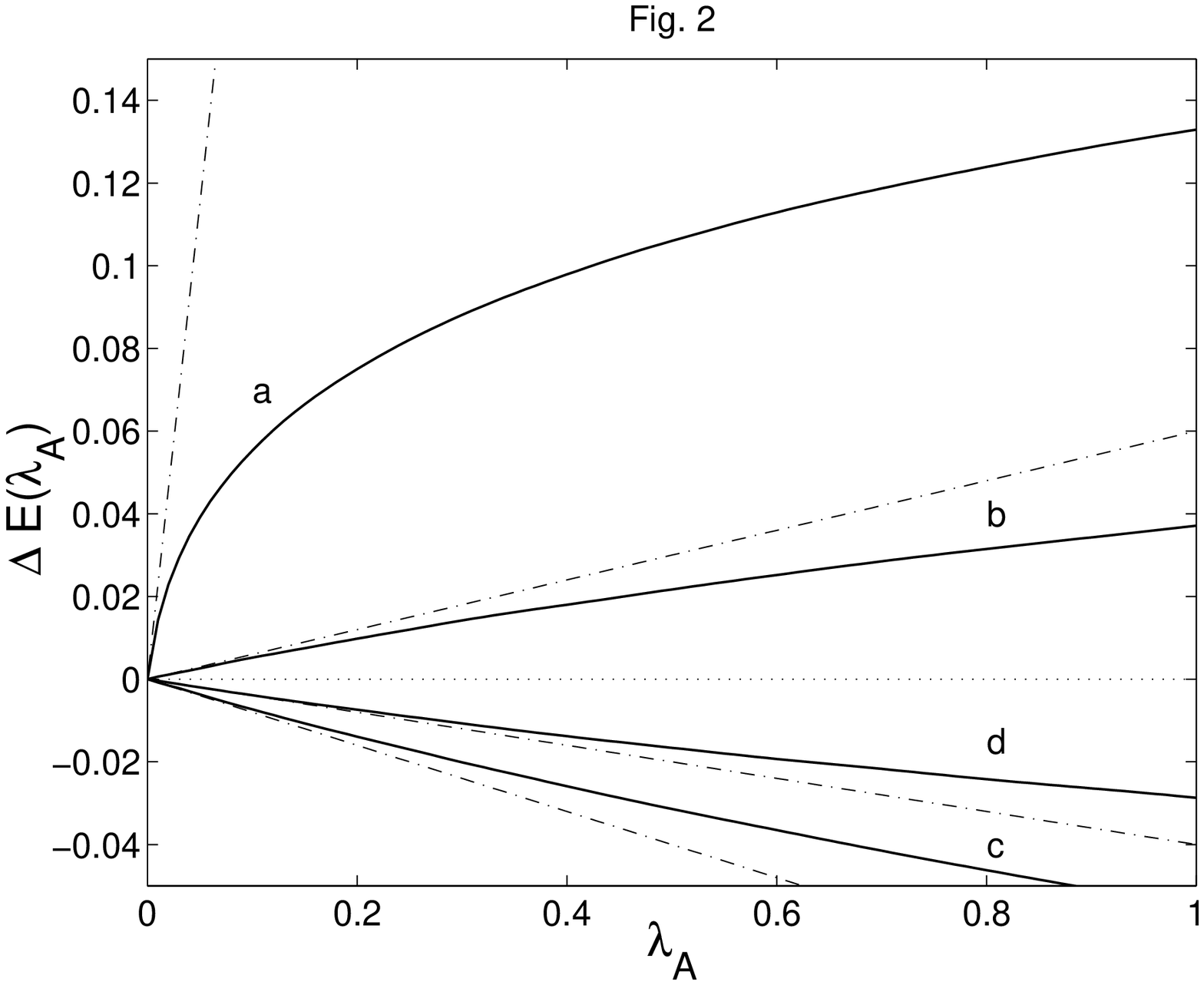}}\\
\end{figure}

\clearpage
\newpage

\begin{figure}
\centering
\vspace{-1cm}
\mbox{  \epsfysize=12.5cm \epsffile{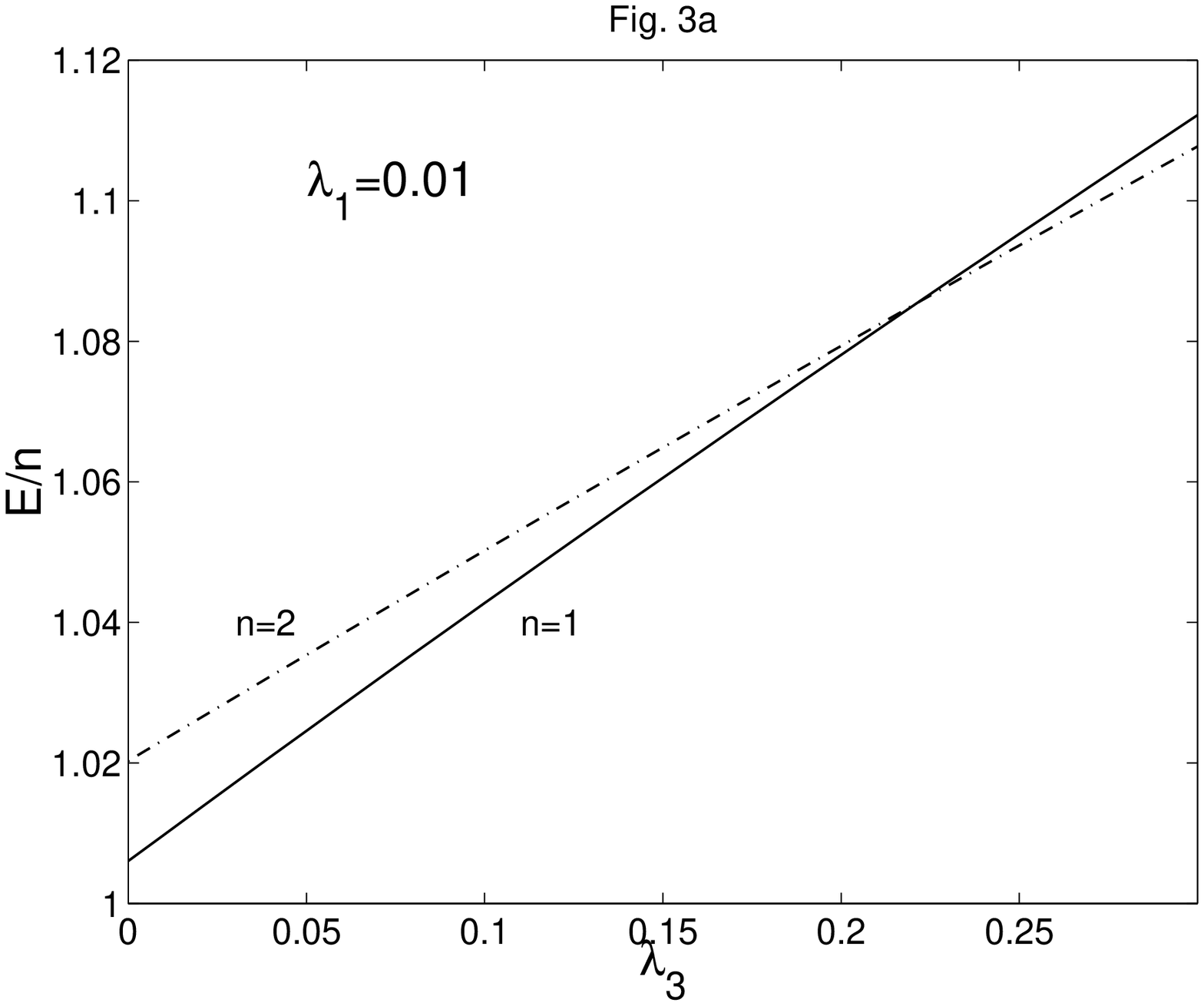}}\\
\end{figure}

\clearpage
\newpage

\begin{figure}
\centering
\vspace{-1cm}
\mbox{  \epsfysize=12.5cm \epsffile{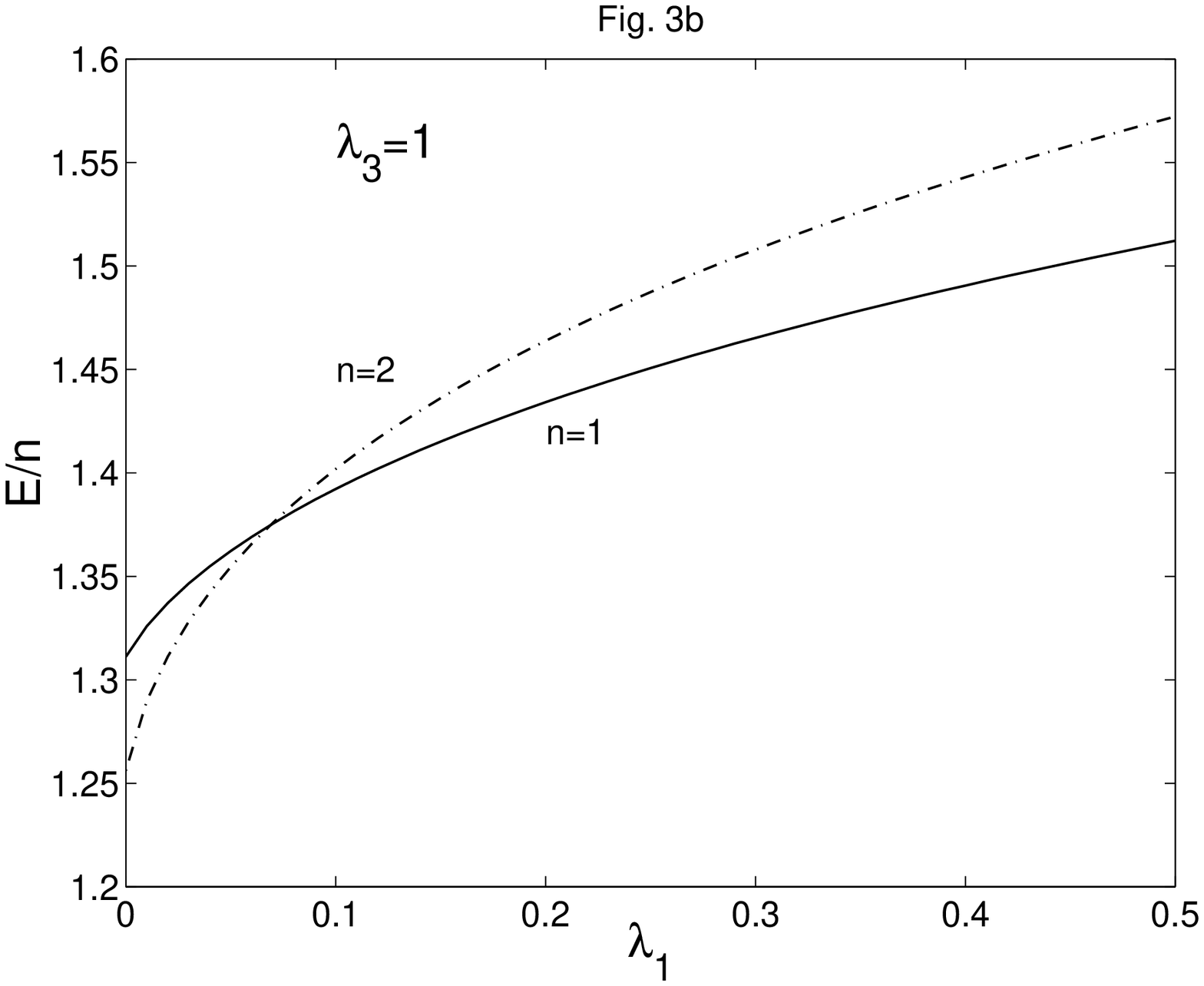}}\\
\end{figure}

\clearpage
\newpage

\begin{figure}
\centering
\vspace{-1cm}
\mbox{  \epsfysize=12.5cm \epsffile{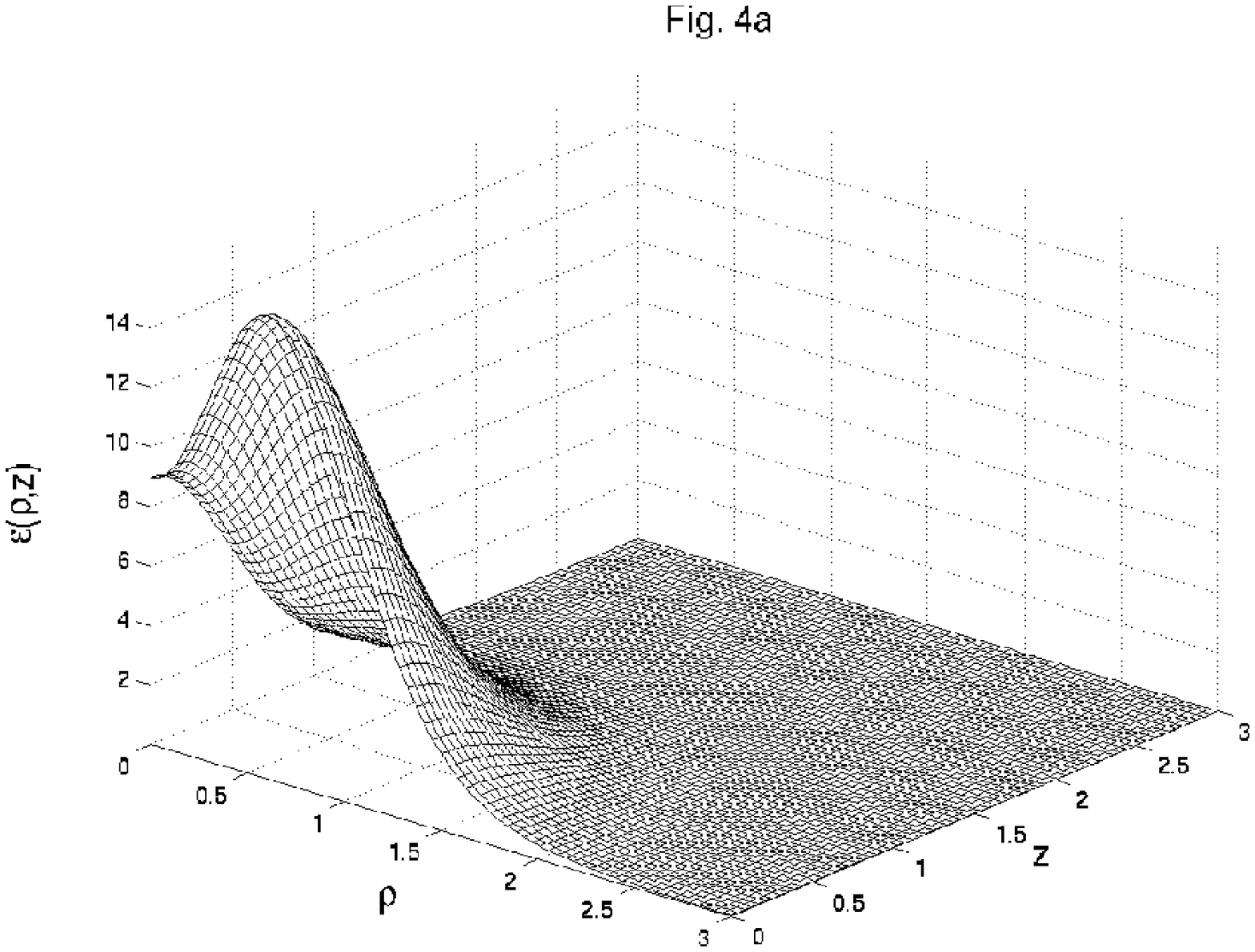}}\\
\end{figure}

\clearpage
\newpage

\begin{figure}
\centering
\vspace{-1cm}
\mbox{  \epsfysize=12.5cm \epsffile{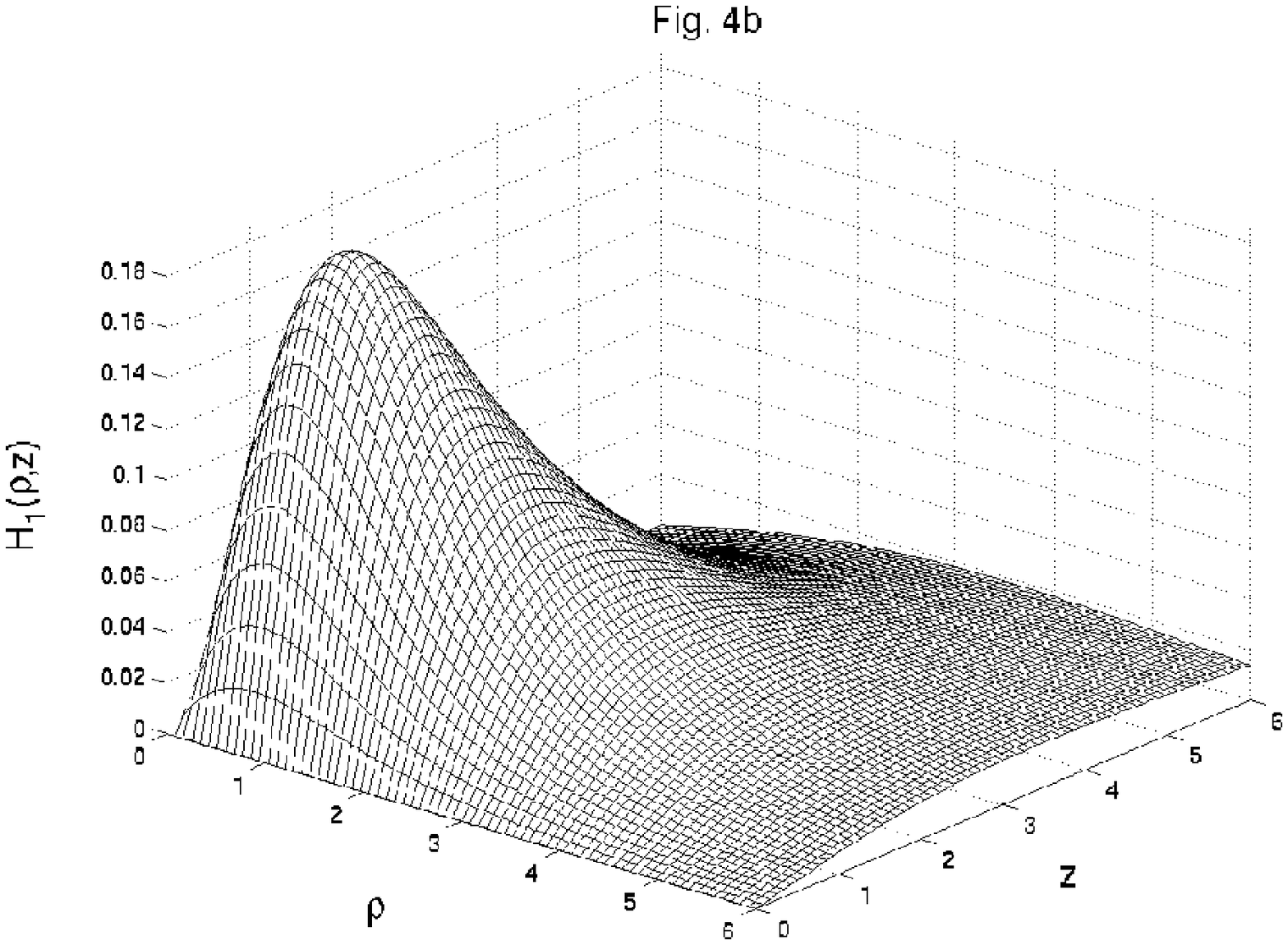}}\\
\end{figure}

\clearpage
\newpage

\begin{figure}
\centering
\vspace{-1cm}
\mbox{  \epsfysize=12.5cm \epsffile{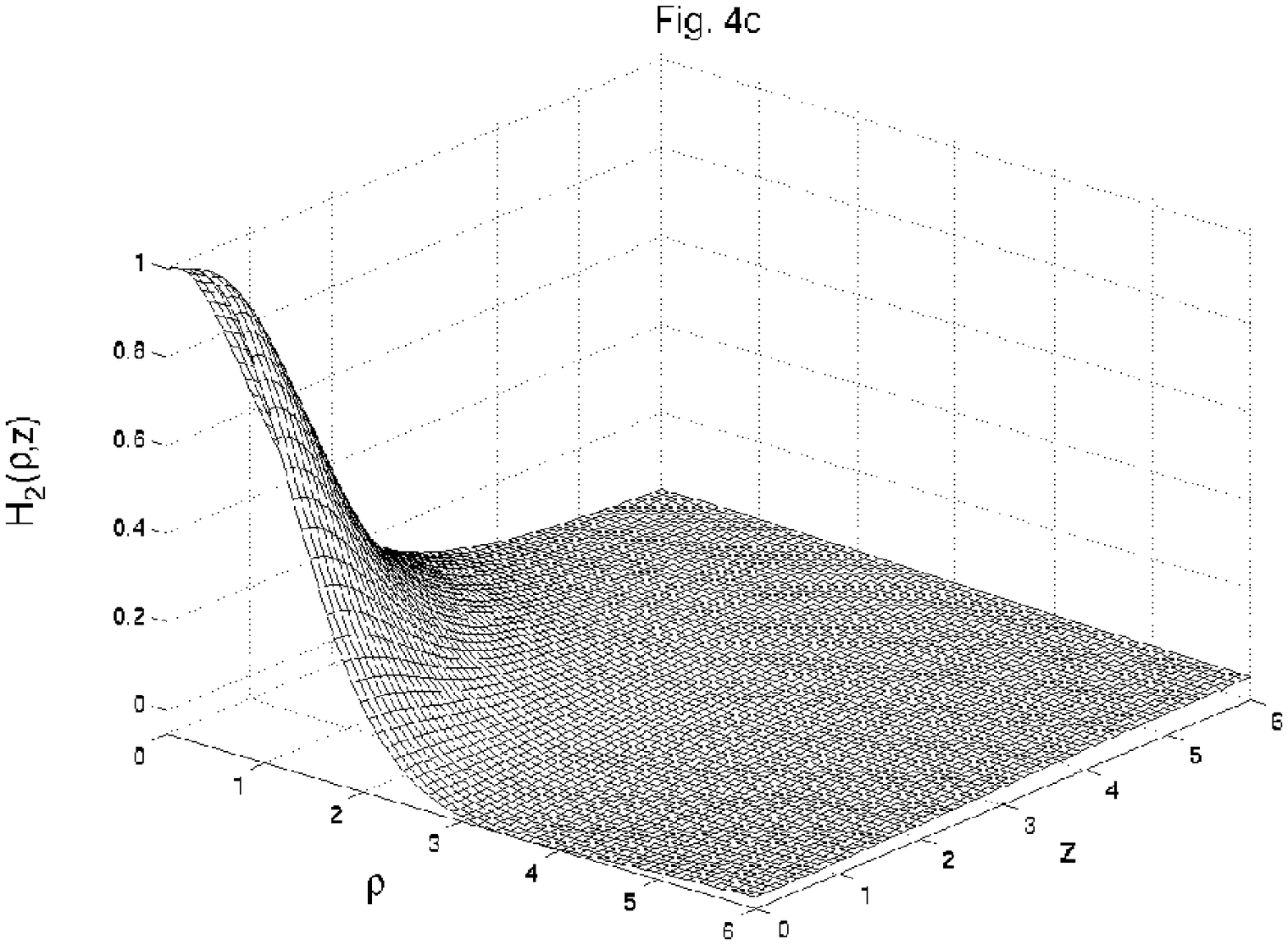}}\\
\end{figure}

\clearpage
\newpage

\begin{figure}
\centering
\vspace{-1cm}
\mbox{  \epsfysize=12.5cm \epsffile{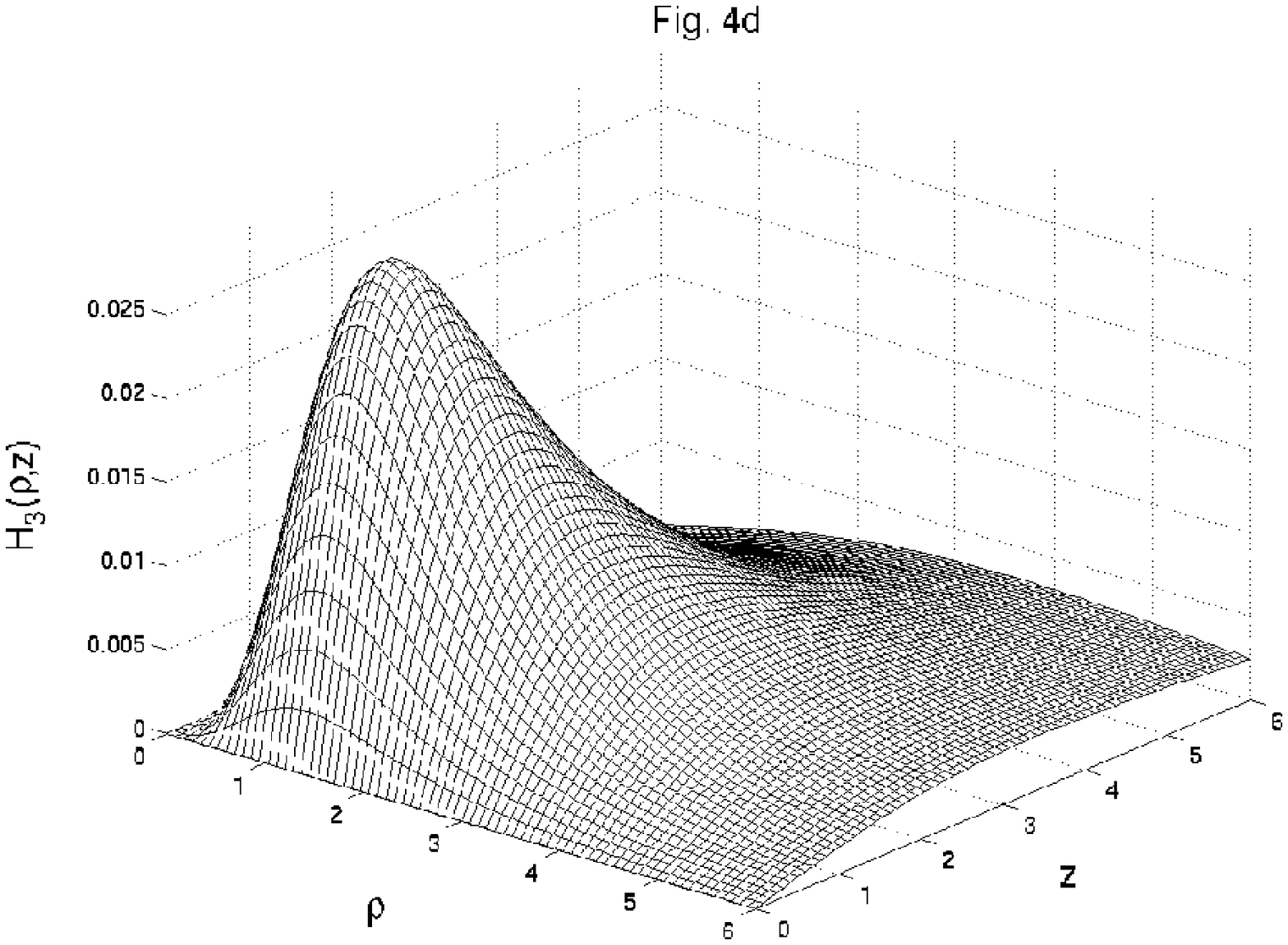}}\\
\end{figure}
\clearpage
\newpage

\begin{figure}
\centering
\vspace{-1cm}
\mbox{  \epsfysize=12.5cm \epsffile{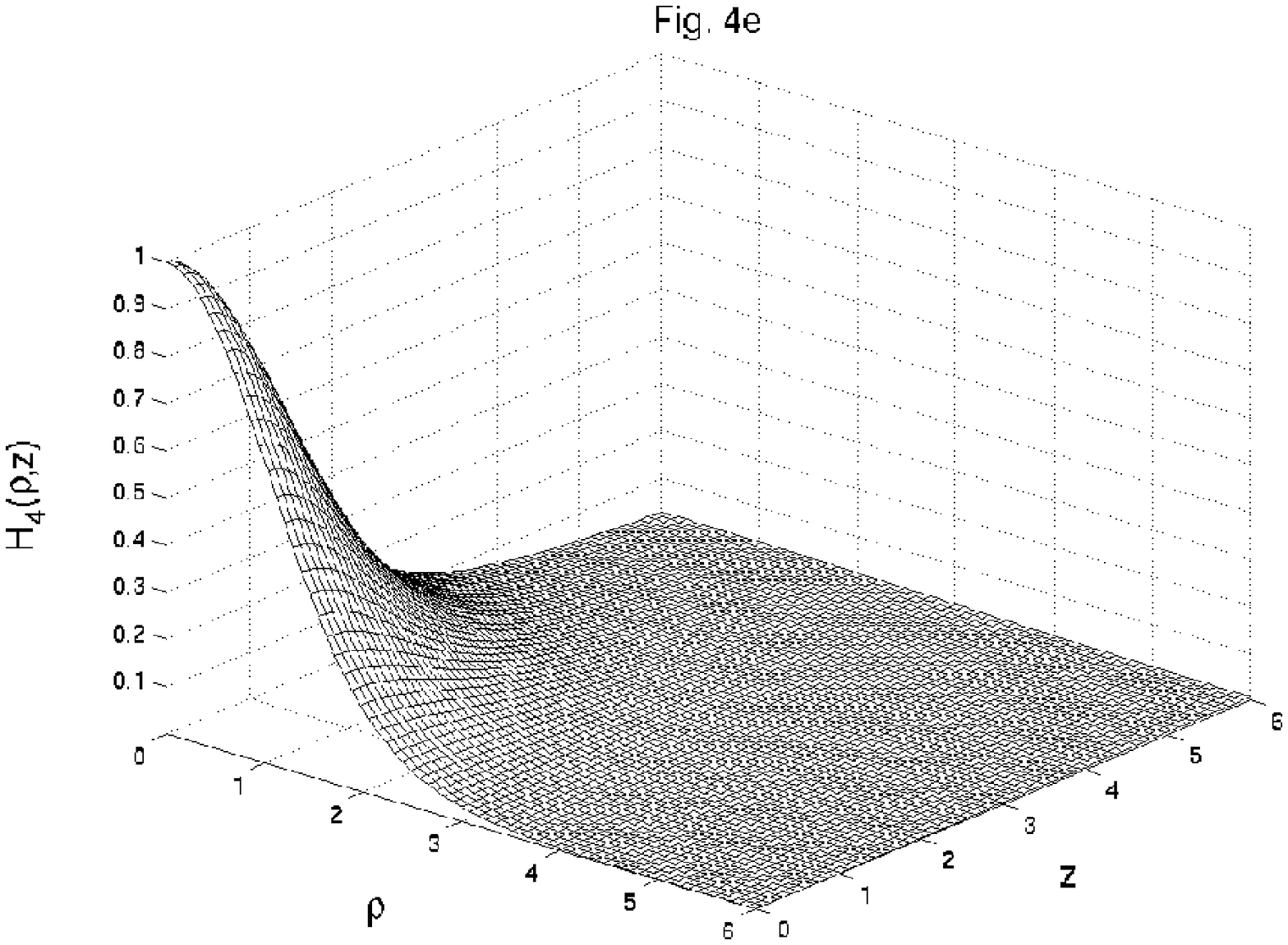}}\\
\end{figure}

\clearpage
\newpage

\begin{figure}
\centering
\vspace{-1cm}
\mbox{  \epsfysize=12.5cm \epsffile{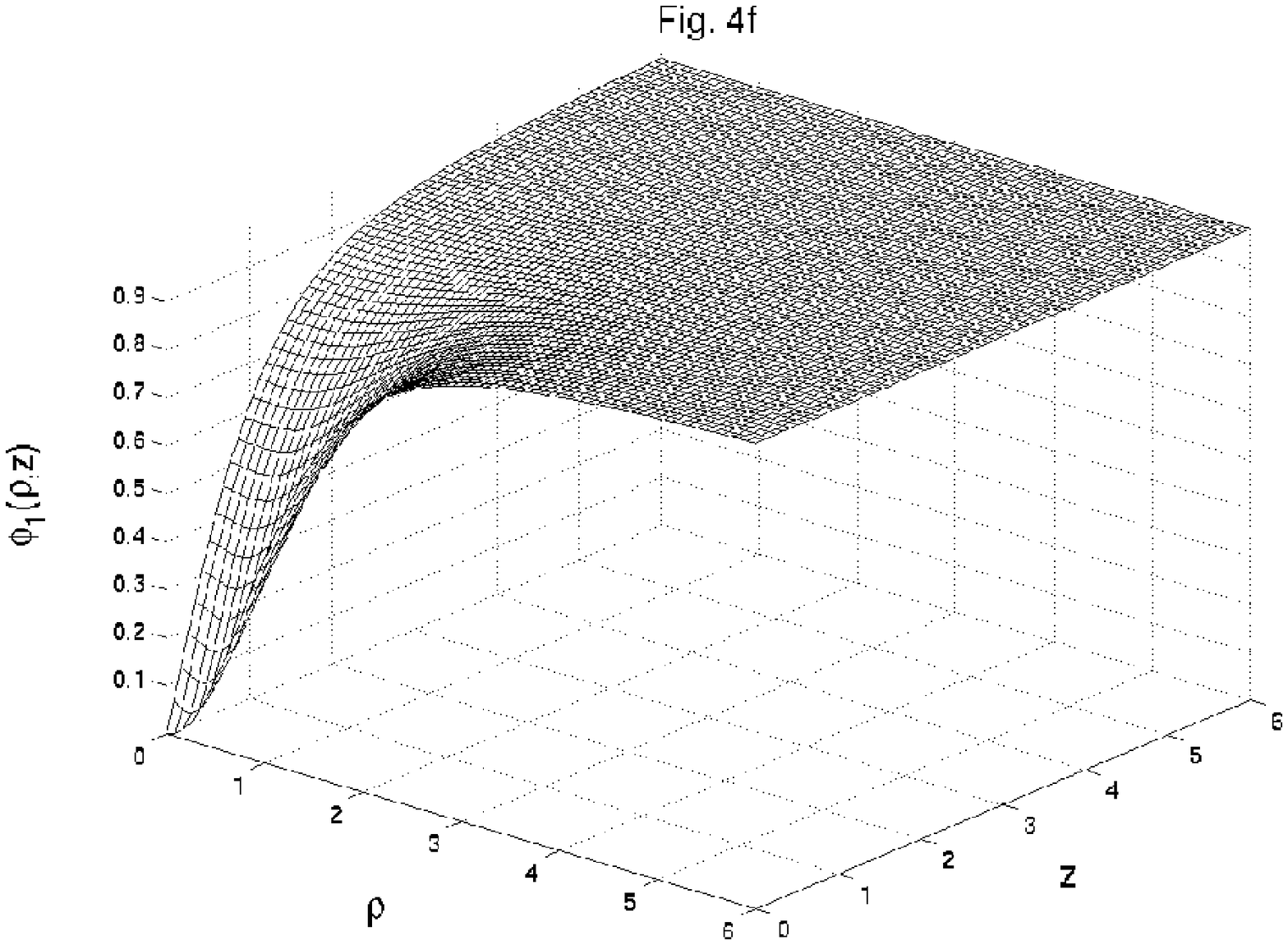}}\\
\end{figure}

\clearpage
\newpage

\begin{figure}
\centering
\vspace{-1cm}
\mbox{  \epsfysize=12.5cm \epsffile{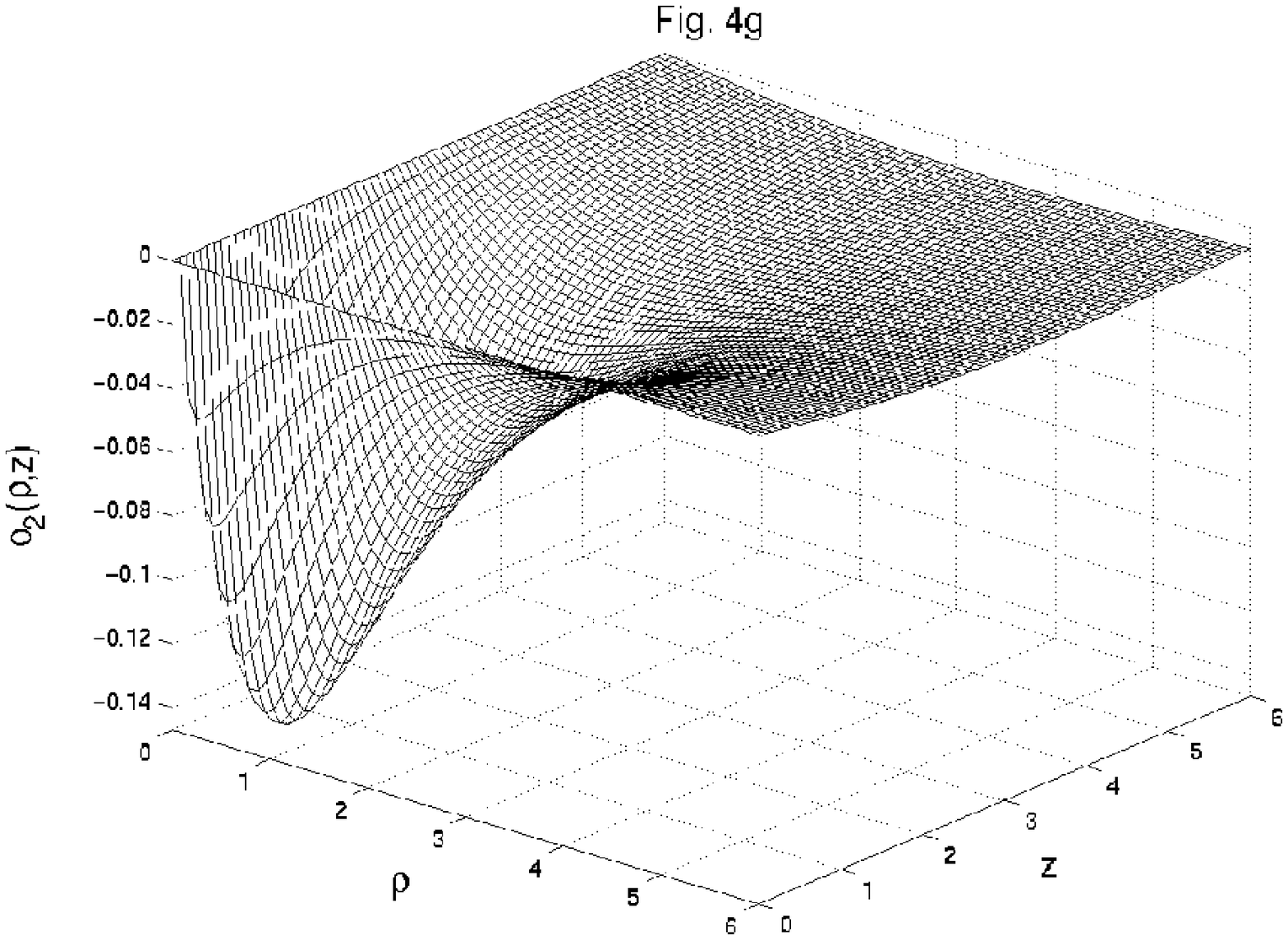}}\\
\end{figure}

\begin{figure}
\centering
\vspace{-1cm}
\mbox{  \epsfysize=12.5cm \epsffile{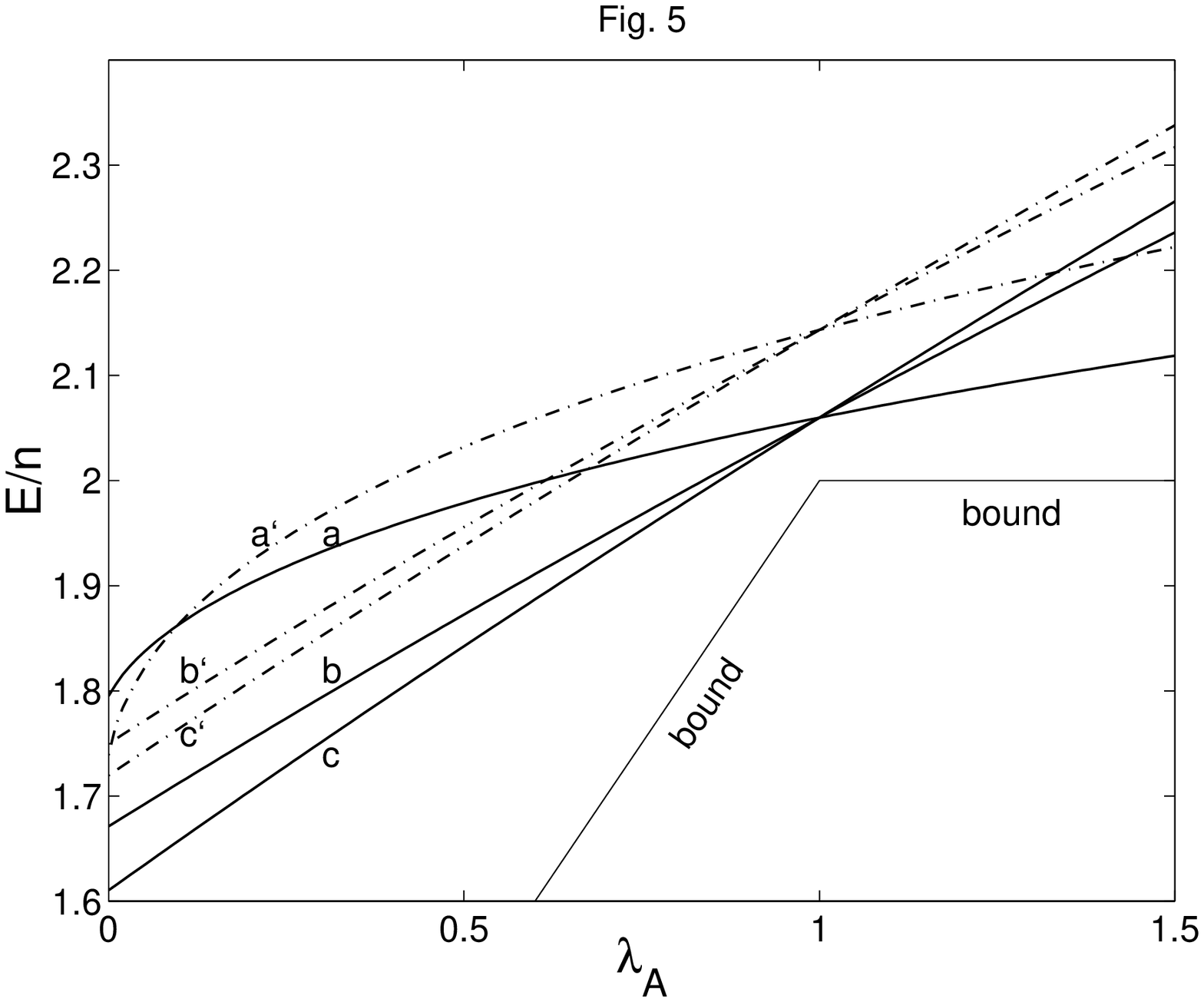}}\\
\end{figure}

\clearpage
\newpage

\begin{figure}
\centering
\vspace{-1cm}
\mbox{  \epsfysize=12.5cm \epsffile{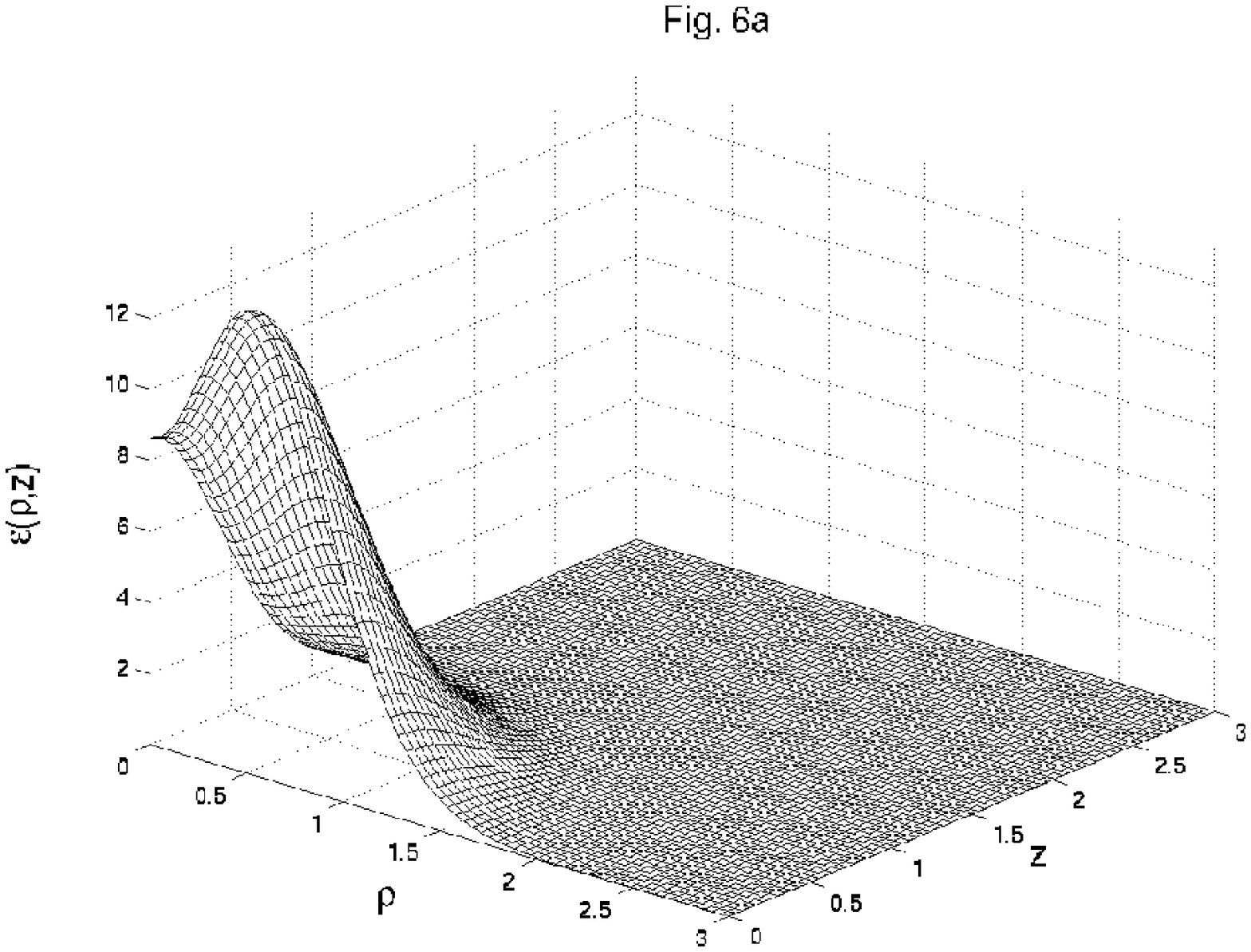}}\\
\end{figure}

\clearpage
\newpage

\begin{figure}
\centering
\vspace{-1cm}
\mbox{  \epsfysize=12.5cm \epsffile{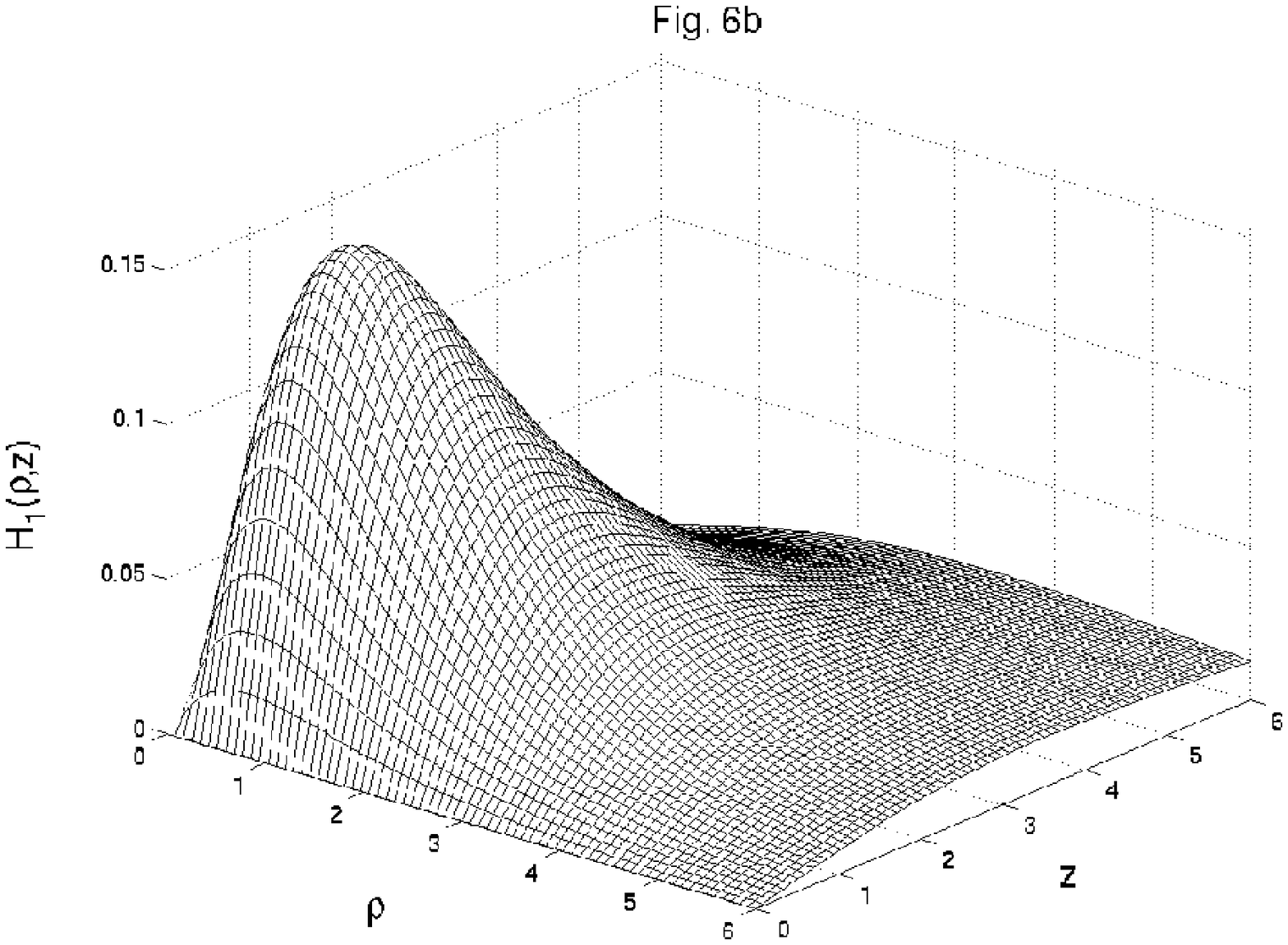}}\\
\end{figure}

\clearpage
\newpage

\begin{figure}
\centering
\vspace{-1cm}
\mbox{  \epsfysize=12.5cm \epsffile{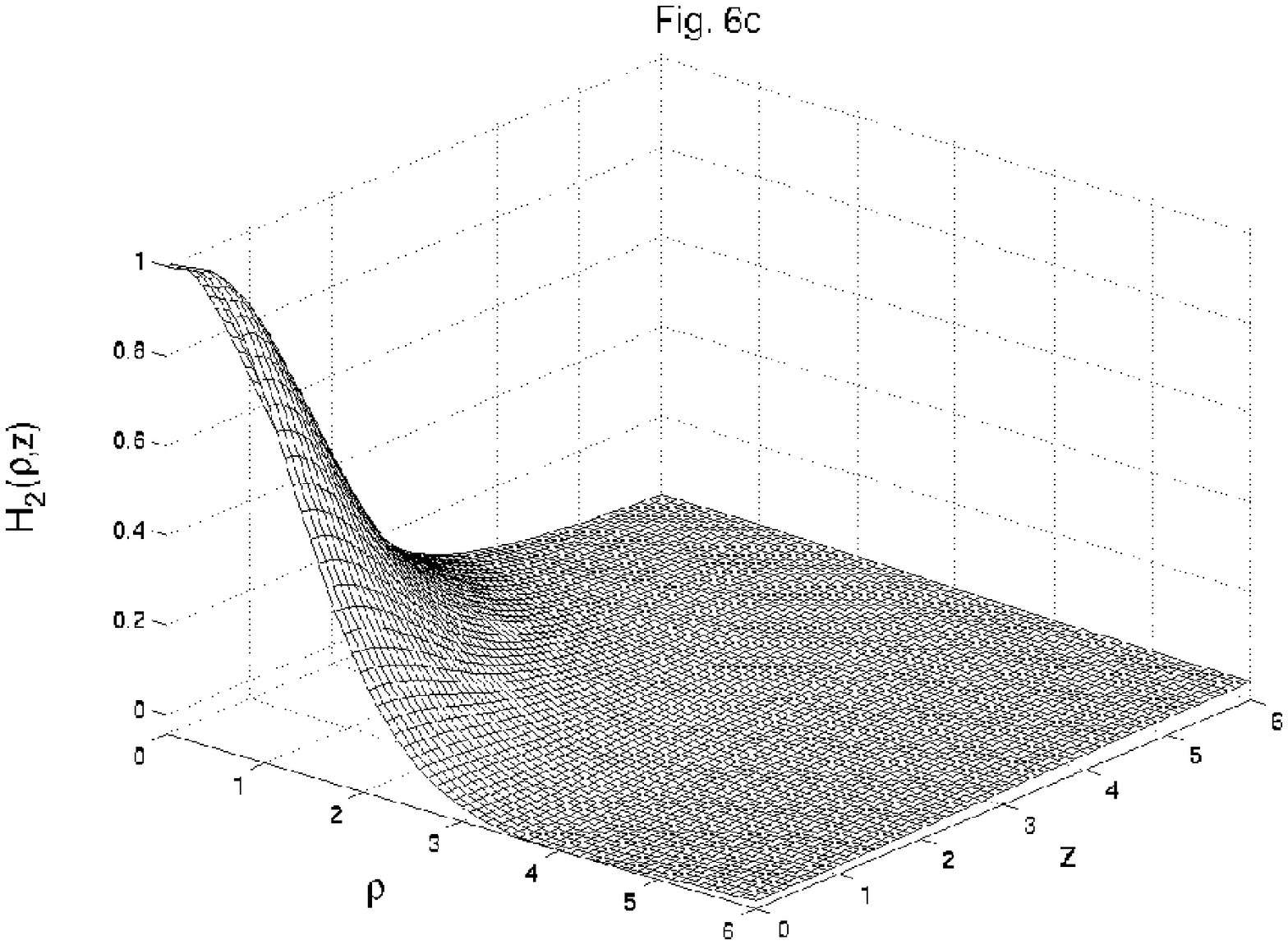}}\\
\end{figure}

\clearpage
\newpage

\begin{figure}
\centering
\vspace{-1cm}
\mbox{  \epsfysize=12.5cm \epsffile{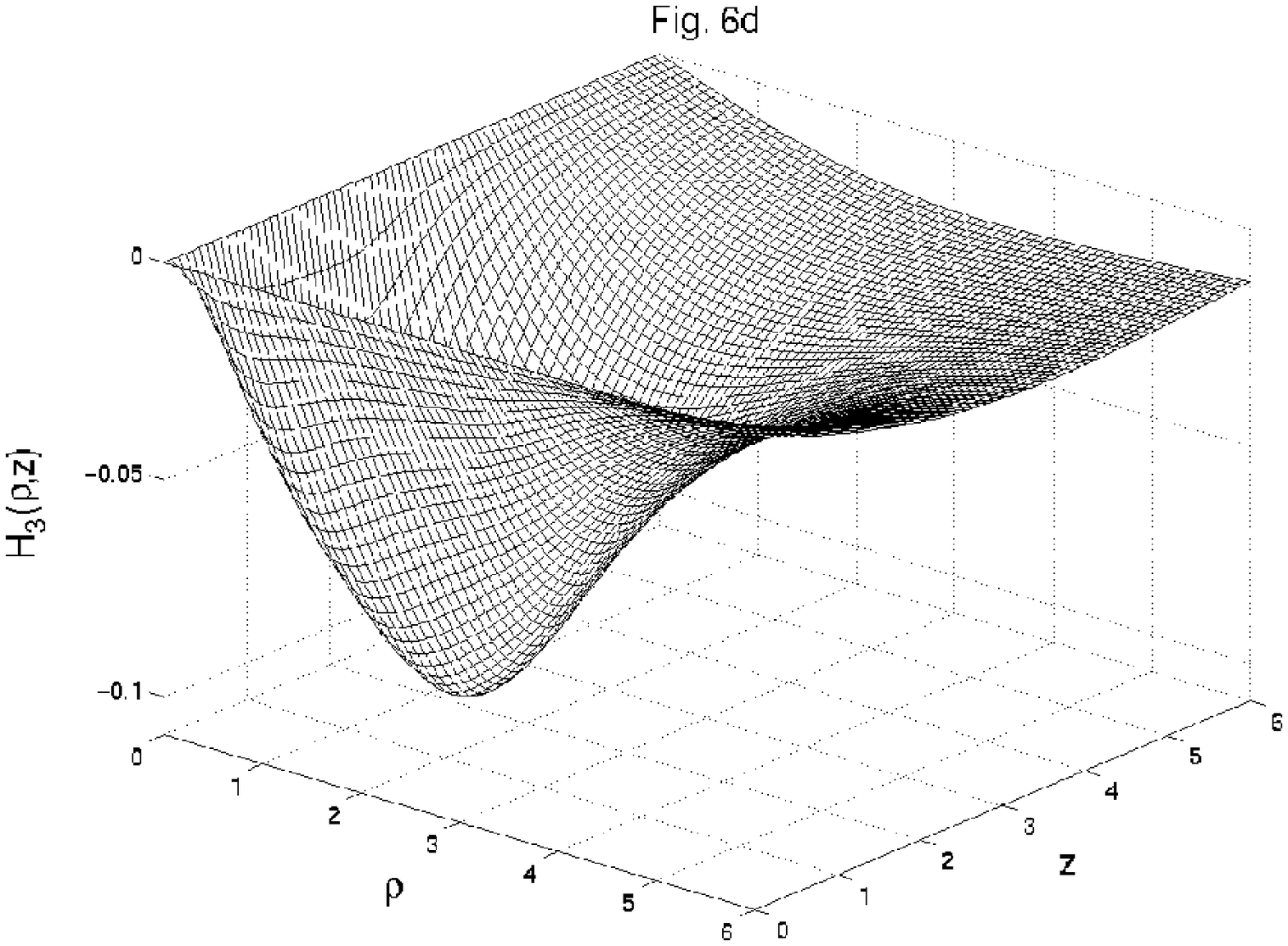}}\\
\end{figure}
\clearpage
\newpage

\begin{figure}
\centering
\vspace{-1cm}
\mbox{  \epsfysize=12.5cm \epsffile{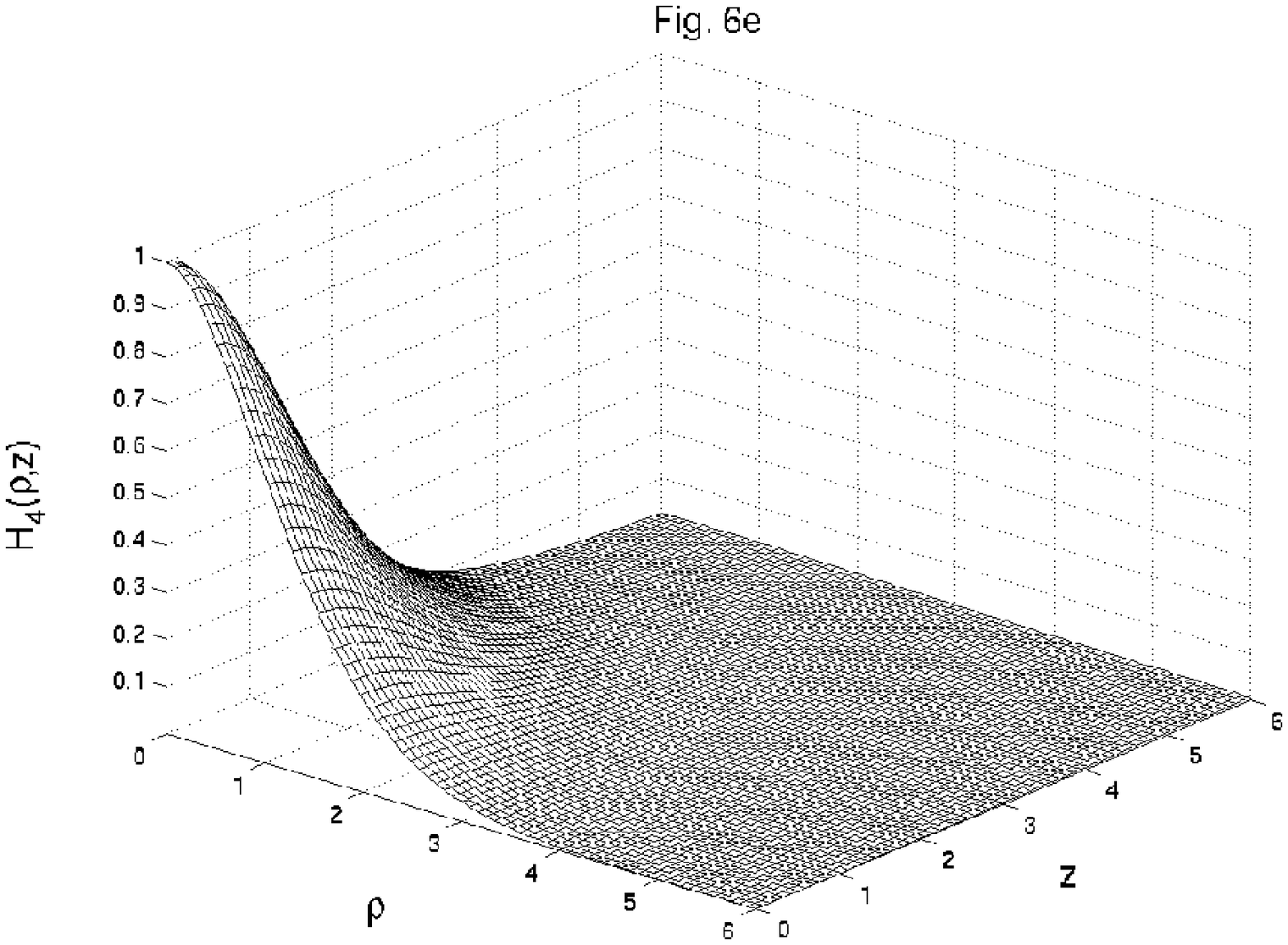}}\\
\end{figure}

\clearpage
\newpage

\begin{figure}
\centering
\vspace{-1cm}
\mbox{  \epsfysize=12.5cm \epsffile{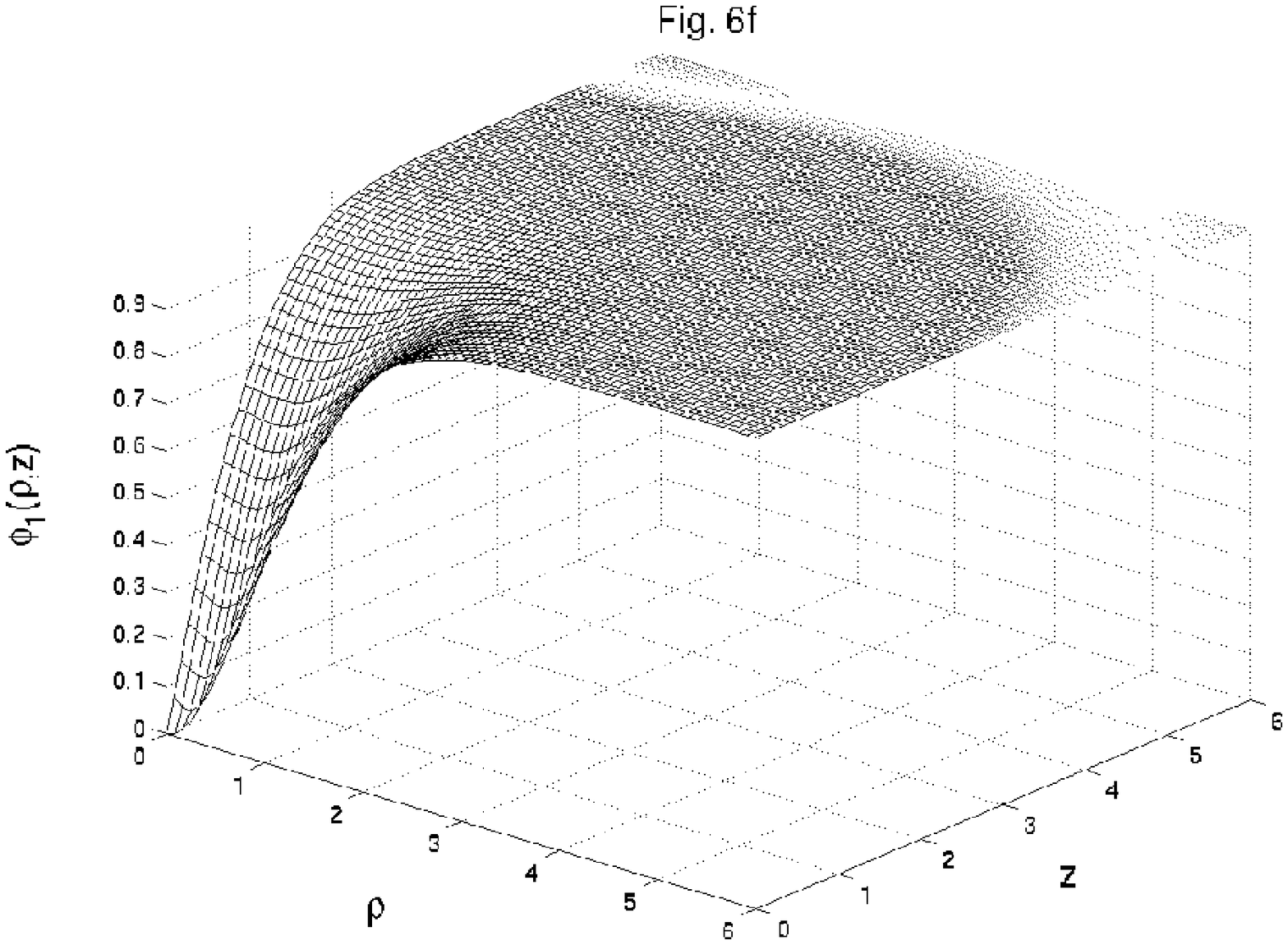}}\\
\end{figure}

\clearpage
\newpage

\begin{figure}
\centering
\vspace{-1cm}
\mbox{  \epsfysize=12.5cm \epsffile{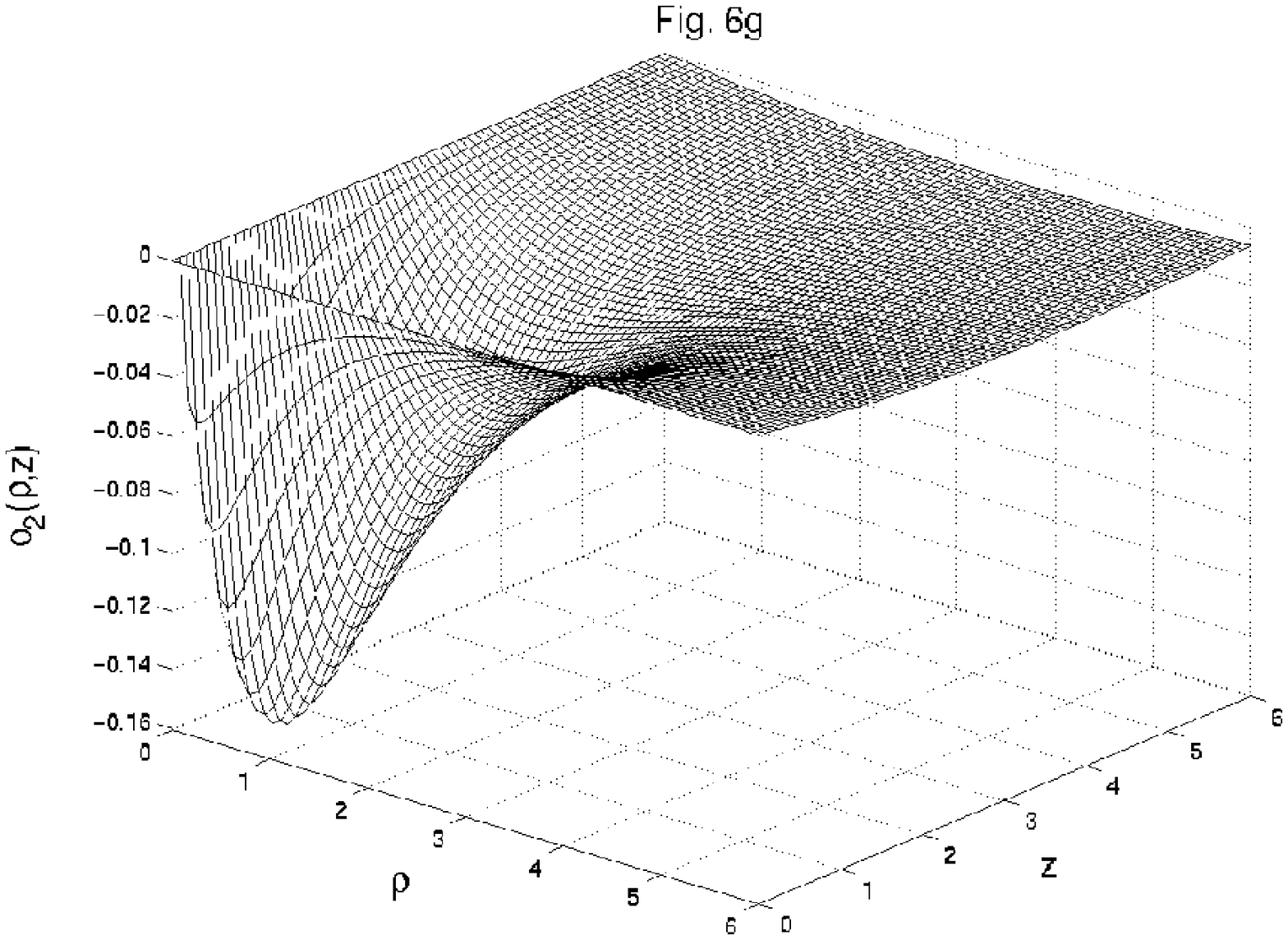}}\\
\end{figure}

\clearpage
\newpage

\begin{figure}
\centering
\vspace{-1cm}
\mbox{  \epsfysize=12.5cm \epsffile{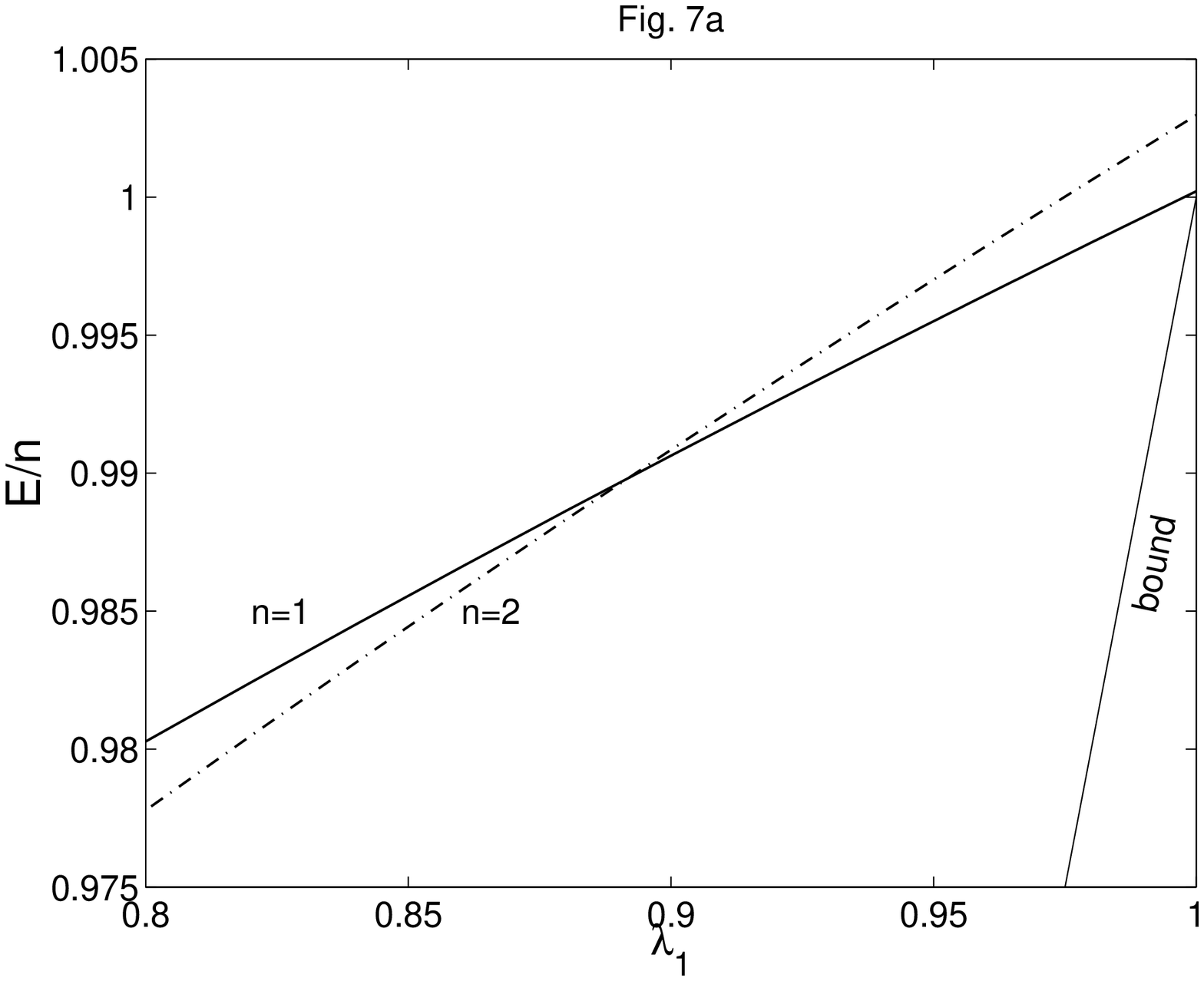}}\\
\end{figure}

\clearpage
\newpage

\begin{figure}
\centering
\vspace{-1cm}
\mbox{  \epsfysize=12.5cm \epsffile{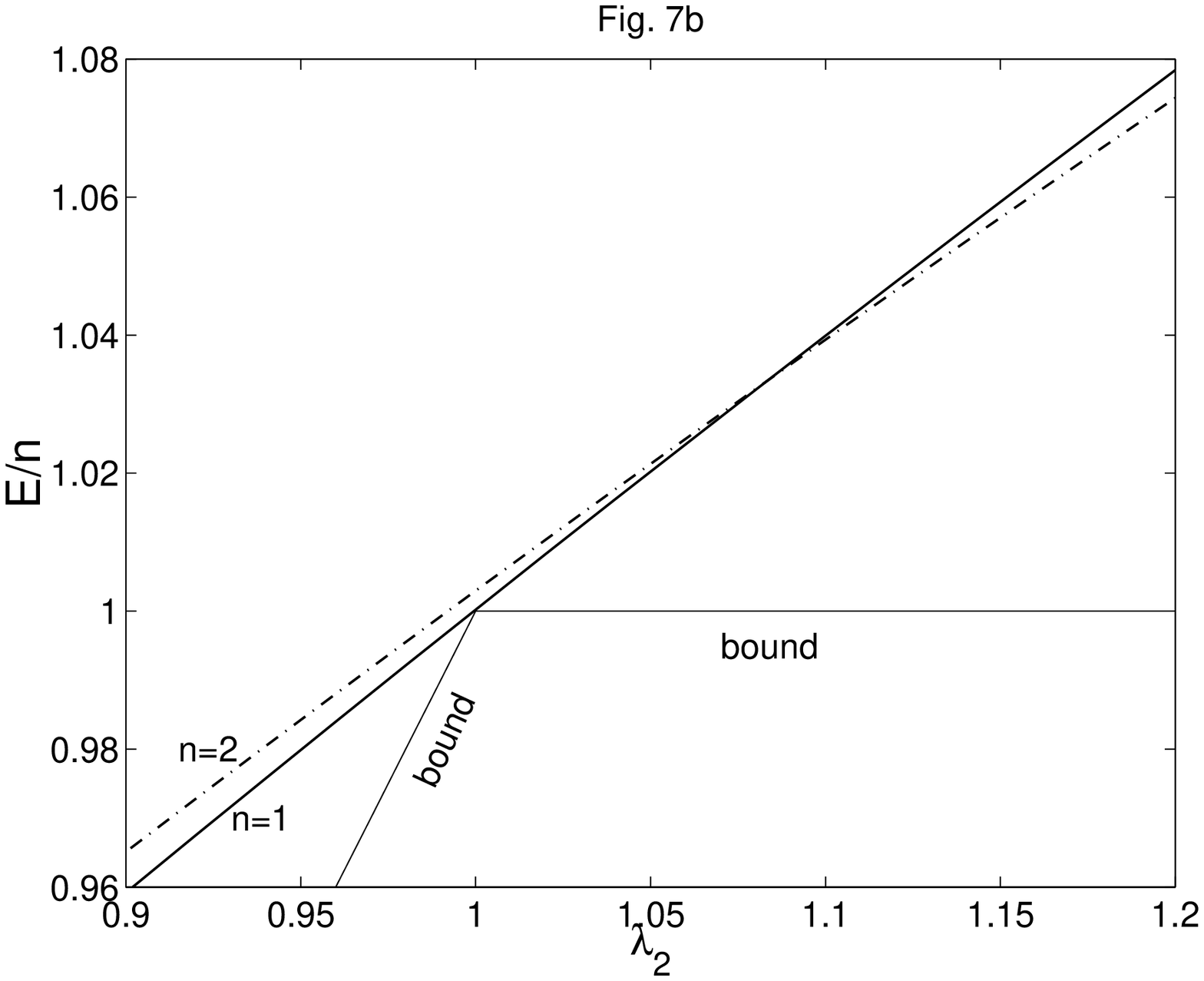}}\\
\end{figure}

\clearpage
\newpage

\begin{figure}
\centering
\vspace{-1cm}
\mbox{  \epsfysize=12.5cm \epsffile{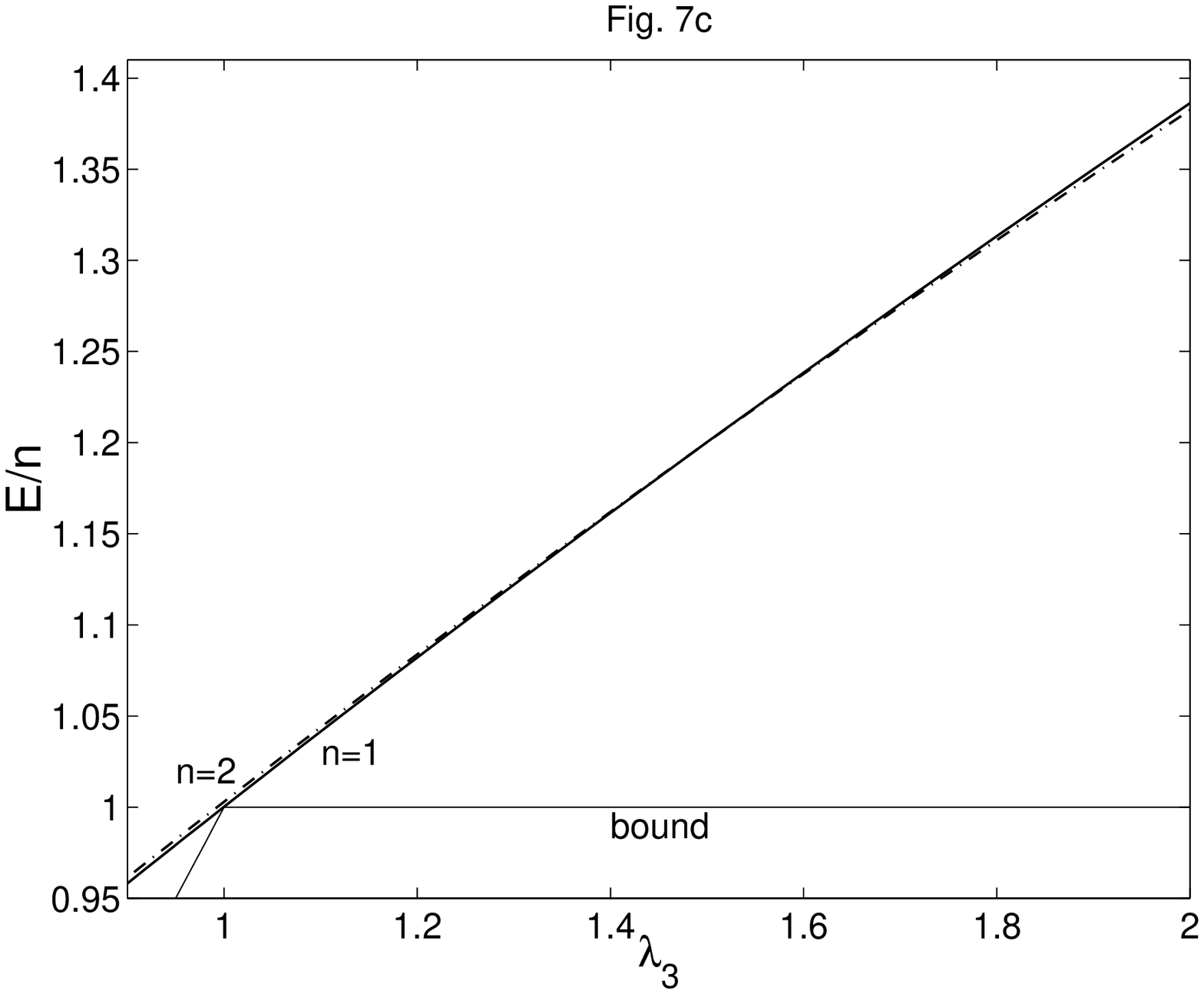}}\\
\end{figure}

\end{document}